\pdfoutput=1

\documentclass[twocolumn,final]{svjour3}

\smartqed

\usepackage{graphicx,amssymb}
\usepackage{epsf,psfig}
\usepackage{txfonts}
\usepackage{natbib}

\journalname{Nonlinear Dynamics}

\begin{document}

\title{Revealing the evolution, the stability and the escapes of families of resonant periodic orbits in Hamiltonian systems}

\author{Euaggelos E. Zotos}

\institute{Euaggelos E. Zotos:
\at Department of Physics, School of Science, \\
Aristotle University of Thessaloniki \\
GR-541 24, Thessaloniki, Greece \\
email:{evzotos@physics.auth.gr}
}

\date{Received: 26 January 2013 / Accepted: 21 February 2012}

\titlerunning{Computing families of resonant periodic orbits in Hamiltonian systems}

\authorrunning{Euaggelos E. Zotos}

\maketitle

\begin{abstract}

We investigate the evolution of families of periodic orbits in a bisymmetrical potential made up of a two-dimensional harmonic oscillator with only one quartic perturbing term, in a number of resonant cases. Our main objective is to compute sufficiently and accurately the position and the period of the periodic orbits. For the derivation of the above quantities (position and period) we deploy in each resonance case semi-numerical methods. The comparison of our semi-numerical results with those obtained by the numerical integration of the equations of motion indicates that, in every case the relative error is always less than 1\% and therefore, the agreement is more than sufficient. Thus, we claim that semi-numerical methods are very effective tools for computing periodic orbits. We also study in detail, the case when the energy of the orbits is larger than the escape energy. In this case, the periodic orbits in almost all resonance families become unstable and eventually escape from the system. Our target is to calculate the escape period and the escape position of the periodic orbits and also monitor their evolution with respect to the value of the energy.

\keywords{Numerical methods; Hamiltonian systems; periodic orbits; escapes}

\end{abstract}

\section{Introduction}
\label{intro}

Periodic orbits are present almost always in the study of differential equations. Keplerian motion under an inverse square law force is elliptic and periodic and the solution of the three-body problem that forms the foundation of the Hill-Brown
theory in celestial mechanics is also periodic. Every area of science has its own oscillatory phenomena and these are usually periodic solutions of differential equations or systems composed of differential equations.

Obtaining periodic orbits in dynamical systems is the main source of information about how the orbits in general are organized and, in Poincar\'{e}'s words, they offer ``the only opening through which we might try to penetrate the fortress (chaos) which has the reputation of being impregnable". The stable periodic orbits explain the dynamics of bounded regular motion, while the unstable periodic orbits in a chaotic set (attractor or saddle) determine its structure [\citealp{24}] and thus, the chaotic behavior of the dynamical system. On this basis, there is no doubt, that periodic orbits represent the backbone of the entire set of orbits and therefore, play a very important role in understanding the orbital structure in a dynamical system. Thus, it is not surprising the high number of works that use the detection of periodic orbits in the analysis of dynamical systems in many different fields. The analysis and control of chaotic dynamical systems [\citealp{36}], the ``scar" theory in quantum mechanics [\citealp{45}], electrical and magnetic fields [\citealp{34}], hydrodynamical flows [\citealp{33}], electrical circuits and optical systems [\citealp{1}] and celestial mechanics [\citealp{29},\citealp{37},\citealp{44},\citealp{50}] are only a few examples.

In an effort to understand the structure of the solutions of non-integrable dynamical systems, numerical determination of their periodic solutions and their stability properties plays a role of fundamental importance. The fact that for most dynamical systems the periodic orbits are dense in the set of all possible solutions, at least in certain parts of the phase space, necessitates the presence of an efficient numerical method for their determination. Over the last decades, several works developing different numerical algorithms in order to compute periodic orbits have been presented. The use of differential correction algorithms for the numerical computation of either two or three dimensional periodic orbits is not a new result. The contributions [\citealp{6},\citealp{21},\citealp{27}] in reference to systems with two degrees of freedom, or [\citealp{4},\citealp{16},\citealp{28},\citealp{32},\citealp{39}] with respect to systems with three degrees of freedom, can be mentioned among many others.

Usually, a dynamical system admits some kind of symmetry, and consequently the traditional approaches used for computing periodic orbits were based on those symmetries. However, when dealing with force fields without symmetries, a different approach must be used. The normal procedure is then the application of the Poincar\'{e} map, and differential corrections are obtained through the computation of the state transition matrix along the periodic orbit. On the other hand, for conservative systems the monodromy matrix has one unit eigenvalue with multiplicity two - related to the time invariance of the system - thus preventing the computation of the corrections. Two approaches are normally used to compute the nontrivial eigenvalues, both of them based on the integration of the variations in Cartesian coordinates. The first computes the complete state transition matrix and uses basic techniques of matrix algebra for obtaining the eigenvalues of a singular matrix. The second eliminates the two unit eigenvalues from the system by creating a lower dimensional map.

In this work, we shall study the evolution of different types of families of resonant periodic orbits. For values of energy larger than the energy of escape the surface of section is partitioned in different escape regions which are defined by the intersections of the asymptotic manifolds of the Lyapunov orbits with the surface of section. In other words, the equipotential curves or, equivalently, the zero-velocity curves (ZVCs) are open containing several channels through which a particle can escape from the dynamical system. The phenomenon of escapes from a dynamical system, especially the escape of stars from stellar systems has been an active field of research during the last decades [\citealp{13},\citealp{15},\citealp{17}-\citealp{19},\citealp{23},\citealp{30},\citealp{40},\citealp{41},
\citealp{46}]. The reader can find more illuminating details on the subject of escapes in the review [\citealp{42}] and also in [\citealp{14}].

The present paper is organized as follows. In the next Section, we describe the properties of the Hamiltonian system that we use in our quest for families of periodic orbits. In the same Section, we provide some theoretical information regarding the location and the stability of the periodic orbits. In Section \ref{famorb}, we present in detail our main results investigating each resonance case separately. In the following Section, we study escaping orbits when $h > h_{esc}$. The paper ends with Section \ref{disc}, which contains the discussion with some concluding remarks.

\section{Presentation of the Hamiltonian system and theoretical background}
\label{systheor}

Over the last half century, dynamical systems made up of harmonic oscillators have been extensively used extensively, in order to describe local motion (i.e near an equilibrium point) [\citealp{2},\citealp{5},\citealp{8},\citealp{9},\citealp{25},\citealp{26},\citealp{38},\citealp{47},\citealp{49}]. In order to study these systems, scientists have used either numerical [\citealp{10},\citealp{31},\citealp{48}] or analytical methods [\citealp{7},\citealp{20},\citealp{22}]. Astronomers frequently use potentials made up of harmonic oscillators, in order to study local motion in galaxies. Of particular interest, are the bisymmetrical potentials, as these systems have been used in order to describe motion near the central parts of elliptical galaxies. A large part of the above studies have been devoted to locate the position and calculate the period of the periodic orbits.

The general form of a two-dimensional dynamical system composed of perturbed harmonic oscillators is
\begin{equation}
V = \frac{1}{2}\left(\omega_1^2 x^2 + \omega_2^2 y^2 \right) + \varepsilon V_1,
\label{genform}
\end{equation}
where $\omega_1$ and $\omega_2$ are the unperturbed frequencies of oscillations along the $x$ and $y$ axes respectively, $\varepsilon$ is the perturbation parameter, while $V_1$ is the function containing the perturbing terms. This is called a perturbed elliptic oscillator.

In the present article we shall try to obtain the starting position and the period of periodic orbits in the general resonance case $\omega_1:\omega_2$ = $n:m$, where of course $n$ and $m$ are integers. We shall use a simple dynamical system with only one quartic perturbing term. The corresponding potential is
\begin{equation}
V = \frac{1}{2}\left(\omega_1^2 x^2 + \omega_2^2 y^2 \right) - \varepsilon x^2y^2.
\label{pot}
\end{equation}
This potential may represent the central parts of deformed galaxies. Moreover, it has been used extensively in calculations of periodic and nonperiodic orbits [\citealp{3},\citealp{12}].

The Hamiltonian to potential (\ref{pot}) is
\begin{equation}
H = \frac{1}{2}\left(\dot{x}^2 + \dot{y}^2 + \omega_1^2 x^2 + \omega_2^2 y^2 \right) - \varepsilon x^2y^2  = h,
\label{ham}
\end{equation}
where $\dot{x}$ and $\dot{y}$ are the momenta per unit mass conjugate to $x$ and $y$, while $h > 0$ is the numerical value of the Hamiltonian, which is conserved.

The equations of motion for a test particle with a unit mass are
\begin{eqnarray}
\ddot{x} &=& - \left[\omega_1^2 - 2\varepsilon y^2 \right] x = - w_1^2 x, \nonumber \\
\ddot{y} &=& - \left[\omega_2^2 - 2\varepsilon x^2 \right] y = - w_2^2 y,
\label{eqmot}
\end{eqnarray}
where, as usual, the dot indicates derivative with respect to the time, while the quantities inside the brackets are considered as the squares of the frequencies $w_1$ and $w_2$ of the oscillations along the $x$ and $y$ axes respectively. Furthermore, the equations governing the evolution of a deviation vector $\vec{\xi} = (\delta x, \delta y, \delta \dot{x}, \delta \dot{y})$ are
\begin{eqnarray}
\dot{(\delta x)} &=& \delta \dot{x}, \nonumber \\
\dot{(\delta y)} &=& \delta \dot{y}, \nonumber \\
\dot{(\delta \dot{x})} &=& -\frac{\partial^2 V}{\partial x^2}\delta x - \frac{\partial^2V}{\partial x \partial y} \delta y, \nonumber \\
\dot{(\delta \dot{y})} &=& - \frac{\partial^2V}{\partial y \partial x} \delta x - \frac{\partial^2 V}{\partial y^2}\delta y.
\label{variac}
\end{eqnarray}

\begin{figure}[!tH]
\centering
\resizebox{\hsize}{!}{\rotatebox{0}{\includegraphics*{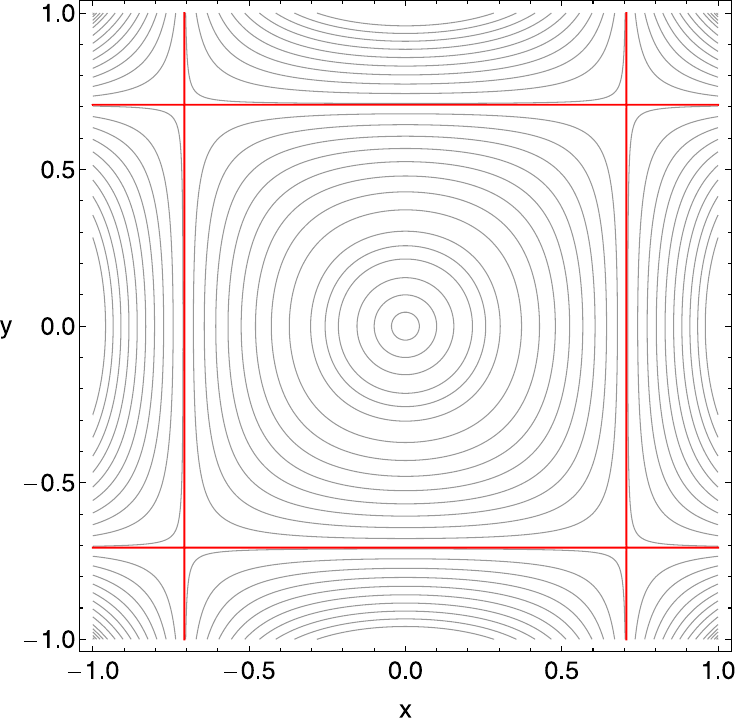}}}
\caption{Equipotential curves of the potential (\ref{pot}) for various values of the energy $h$ when $\omega_1 = \omega_2 = \omega = 1$ and $\varepsilon = 1$. The equipotential curve corresponding to the energy of escape is shown with red color.}
\label{eqpot}
\end{figure}

The equipotential curves of the potential (\ref{pot}) for various values of the energy $h$ are shown in Fig. \ref{eqpot}. Without the loss of generality, we take $\omega_1 = \omega_2 = \omega = 1$ and $\varepsilon = 1$, that is the 1:1 resonance case. The equipotential corresponding to the energy of escape $h_{esc}$ is plotted with red color in the same plot. The energy of escape is given by the following expression
\begin{equation}
h_{esc} = \frac{\omega_1^2\omega_2^2}{4\varepsilon}.
\label{hesc}
\end{equation}
However, if we take into account that
\begin{equation}
\frac{\omega_1}{\omega_2} = \frac{n}{m}
\label{omres}
\end{equation}
and combine Eqs. (\ref{hesc}) and (\ref{omres}) then, we can rewrite the energy of escape as
\begin{equation}
h_{esc} = \frac{m^2}{n^2} \frac{\omega_1^4}{4\varepsilon}.
\label{hesc2}
\end{equation}

In a previous work [\citealp{6}] (hereafter Paper I) some very useful semi-numerical relations for the position and the period of resonant periodic orbits have been introduced. We deem, that it would be appropriate to recall these expressions. To begin with, the frequency of the periodic orbits is given by
\begin{equation}
w_2 = \sqrt{\omega_2^2 - c \frac{m^2}{n^2} \frac{\varepsilon h}{\omega_2^2}}.
\label{freq}
\end{equation}
We observe, that the frequency of the periodic orbits depends on the unperturbed frequency $\omega_2$, the energy $h$, the perturbation parameter $\varepsilon$ and the values $n$ and $m$ defining the particular resonance. However, in Eq. (\ref{freq}) there is an extra term $c$ which is not included in the original expression of Paper I (Eq. (15)). This quantity is a correction term which will allow us to manipulate this basic expression in every resonant case in order to obtain much more reliable results. For periodic orbits starting perpendicularly from the $x$-axis the starting point can be obtained as
\begin{equation}
x_s = \frac{m}{n} \frac{1}{w_2} \sqrt{h},
\label{xs}
\end{equation}
while for periodic orbits going through the origin the starting point is calculated as
\begin{equation}
\dot{x_s} = \sqrt{h}.
\label{pxs}
\end{equation}
The semi-numerical formula regarding the period is the same for both types of orbits and is given by
\begin{equation}
T_s = \frac{2\pi m}{w_2}.
\label{Ts}
\end{equation}

A very interesting and also important issue in the field of periodic orbits is the determination of the stability of a periodic orbit. Let the variational equations of a specific periodic orbit of period $T$ be
\begin{equation}
\dot{\xi_i}(t) = \displaystyle\sum_{j=1}^{4} \frac{\partial F_i}{\partial x_j} \xi_j \ \ \ \ \ \ \ (i,j=1,2,3,4).
\label{varsys}
\end{equation}
If $X(t)$ is the matrix, whose columns are the four solutions of the system (\ref{varsys}) with initial conditions (1,0,0,0), (0,1,0,0), (0,0,1,0) and (0,0,0,1) then, we have the so-called monodromy matrix
\begin{equation}
\displaystyle X(t) = \left(
\begin{array}{cccc}
\frac{\displaystyle \partial x}{\displaystyle \partial x_0} & \frac{\displaystyle \partial x}{\displaystyle \partial y_0} & \frac{\displaystyle \partial x}{\displaystyle \partial \dot{x_0}} & \frac{\displaystyle \partial x}{\displaystyle \partial \dot{y_0}} \\
\frac{\displaystyle \partial y}{\displaystyle \partial x_0} & \frac{\displaystyle \partial y}{\displaystyle \partial y_0} & \frac{\displaystyle \partial y}{\displaystyle \partial \dot{x_0}} & \frac{\displaystyle \partial y}{\displaystyle \partial \dot{y_0}} \\
\frac{\displaystyle \partial \dot{x}}{\displaystyle \partial x_0} & \frac{\displaystyle \partial \dot{x}}{\displaystyle \partial y_0} & \frac{\displaystyle \partial \dot{x}}{\displaystyle \partial \dot{x_0}} & \frac{\displaystyle \partial \dot{x}}{\displaystyle \partial \dot{y_0}} \\
\frac{\displaystyle \partial \dot{y}}{\displaystyle \partial x_0} & \frac{\displaystyle \partial \dot{y}}{\displaystyle \partial y_0} & \frac{\displaystyle \partial \dot{y}}{\displaystyle \partial \dot{x_0}} & \frac{\displaystyle \partial \dot{y}}{\displaystyle \partial \dot{y_0}}
\end{array}
\right).
\label{matrix}
\end{equation}
When $t=T$ there is also a monodromy matrix $X(T)$. The stability of a periodic orbit depends on the eigenvalues of this monodromy matrix. We define the stability index (S.I) as
\begin{equation}
\rm S.I = Tr(X(T)) - 2,
\label{SI}
\end{equation}
where $\rm Tr(X(T)) = \lambda_1 + \lambda_2 + \lambda_3 + \lambda_4$ is the trace of the monodromy matrix, while $\lambda_i$ $(i=1,4)$ are the eigenvalues. Now, according to the value of S.I we can determine if a periodic orbit is stable or unstable. In particular, if $|\rm S.I| < 2$ the periodic orbit is stable, if $|\rm S.I| > 2$ the periodic orbit is unstable, while if $|\rm S.I| = 2$ the periodic orbit is critical stable.

For the numerical integration of the equations of motion and the variational equations, a double precision Bulirsh-St\"{o}er algorithm [\citealp{35}] was used. The accuracy of the calculations was checked by the constancy of the energy integral (\ref{ham}), which was conserved better than one part in $10^{-11}$, although for most orbits it was better than one part in $10^{-12}$.

\section{Families of resonant periodic orbits}
\label{famorb}

In this Section, we will investigate the evolution and the stability of several families of resonant periodic orbits. Our aim is to locate the position of periodic orbits and the corresponding period in the resonant cases using the semi-numerical relations presented in Paper I and then to compare the results with the corresponding outcomes given by the numerical integration of the equations of motion and the variational equations. At the same time, we shall make en effort in order to improve these semi-numerical relations. Here we must point out, that the basic relations regarding the position and the period of the periodic orbits presented in the previous Section, are very simple. Therefore, it would be very difficult to achieve great precision especially in cases where the families of the resonant periodic orbits present either turning points or a non-monotone behavior.

In order to overtake this obstacle, we will exploit the outcomes derived from the numerical integration in each resonance case in an attempt to define very accurate formulas giving the position and the period of the periodic orbits in each family. In particular, taking into account that each family contains 5000 orbits that have been integrated numerically, we will interpolate these data thus obtaining a fourth-order polynomial fitting curve. The main advantage of this procedure is that we can define equations giving the position and the period of the periodic orbits that contain only one variable which is the energy $h$. The fourth-order polynomial fitting curves would be of the following form
\begin{eqnarray}
x_f(h) &=& x_0 + x_1 h + x_2 h^2 + x_3 h^3 + x_4 h^4, \nonumber \\
\dot{x_f}(h) &=& \dot{x_0} + \dot{x_1} h + \dot{x_2} h^2 + \dot{x_3} h^3 + \dot{x_4} h^4, \nonumber \\
T_f(h) &=& T_0 + T_1 h + T_2 h^2 + T_3 h^3 + T_4 h^4,
\label{fitcurs}
\end{eqnarray}
where the values $x_i$, $\dot{x_i}$ and $T_i$ $(i=0, ... , 4)$ are obtained in each resonance case separately by the interpolation of the numerical data. Using this method, we can achieve the best possible agreement between the numerical results obtained from the numerical integration of the equation and the semi-numerical outcomes from the interpolation. In fact, in every resonance case the relative error between numerical and semi-numerical results is always less than 1\%.

Of course, it is not feasible to study all the possible resonance cases. Therefore, we shall investigate only the resonance cases $\omega_1$:$\omega_2$ = 1:1, $\omega_1$:$\omega_2$ = 1:2, $\omega_1$:$\omega_2$ = 1:3, $\omega_1$:$\omega_2$ = 2:3 and $\omega_1$:$\omega_2$ = 3:4. In order to keep things simple, we use the value $\varepsilon = 1$, while the value of the energy $h$, will be treated as a parameter. In particular, in every resonance case we use different increment value regarding the energy $h$. We choose variable energy step $dh$ for every resonance so that the interval $[h_{min}, h_{max}]$ contains always 5000 periodic orbits ($h_{min}$ and $h_{max}$ are the minimum and the maximum value of the energy respectively for which the particular resonance exists). Using this method, we are able to obtain very ``rich" families of periodic orbits which will allow us to establish more accurate fitting curves.

\subsection{The 1:1 resonance}
\label{res11}

\begin{figure}[!tH]
\centering
\resizebox{\hsize}{!}{\rotatebox{0}{\includegraphics*{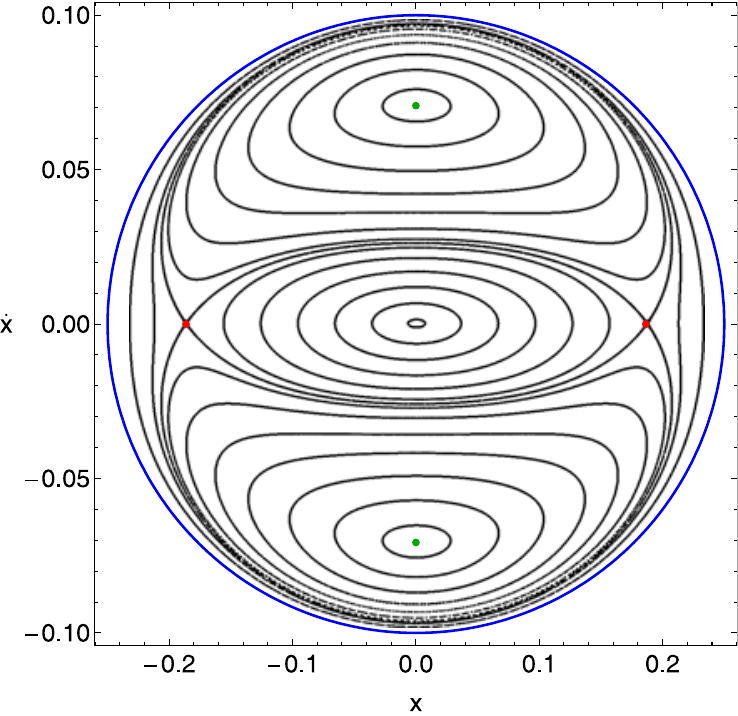}}}
\caption{The $(x, \dot{x})$ phase plane for the 1:1 resonance case. Here, $\omega_1 = \omega_2 = \omega = 0.4$, while the value of the energy is 0.005.}
\label{pss11}
\end{figure}

We shall start from the 1:1 resonance when $\omega_1 = \omega_2 = \omega = 0.4$. Fig. \ref{pss11} shows the $(x, \dot{x})$, $y=0, \dot{y}>0$ Poincar\'{e} Surface of Section (PSS) when $h = 0.005$. We observe, that the entire phase plane is covered only by regular orbits. Apart from the obvious periodic orbit located at the origin $(x_0=\dot{x_0}=0)$, we can identify two additional types of periodic orbits. The first type has two starting positions at the $\dot{x}$-axis which are symmetrical to the $x$-axis. The two green dots in the plot show the exact position of the periodic orbits, while the green color indicate that these are stable periodic orbits. The second type has also two starting positions at the $x$-axis which are symmetrical to the $\dot{x}$-axis, but in this case both points correspond to the same orbit which in fact produces the separatrix. This orbit is unstable and therefore the starting positions are marked with red dots. Using elementary calculations we could prove that all the starting positions for both stable and unstable 1:1 periodic orbits belong to the ellipse $x^2/|x_0|^2 + \dot{x}^2/\dot{|x_0|}^2 = 1$, where $|x_0|$ and $\dot{|x_0|}$ are the starting positions of the unstable and stable periodic orbits at the $x$-axis and $\dot{x}$-axis respectively. The outermost blue line is the Zero Velocity Curve (ZVC) which includes all the invariant curves at the $(x, \dot{x})$ plane and is defined as
\begin{equation}
f_1(x,\dot{x}) = \frac{1}{2}\dot{x} + V(x) = h.
\label{ZVCxpx}
\end{equation}

\begin{figure}[!tH]
\centering
\resizebox{\hsize}{!}{\rotatebox{0}{\includegraphics*{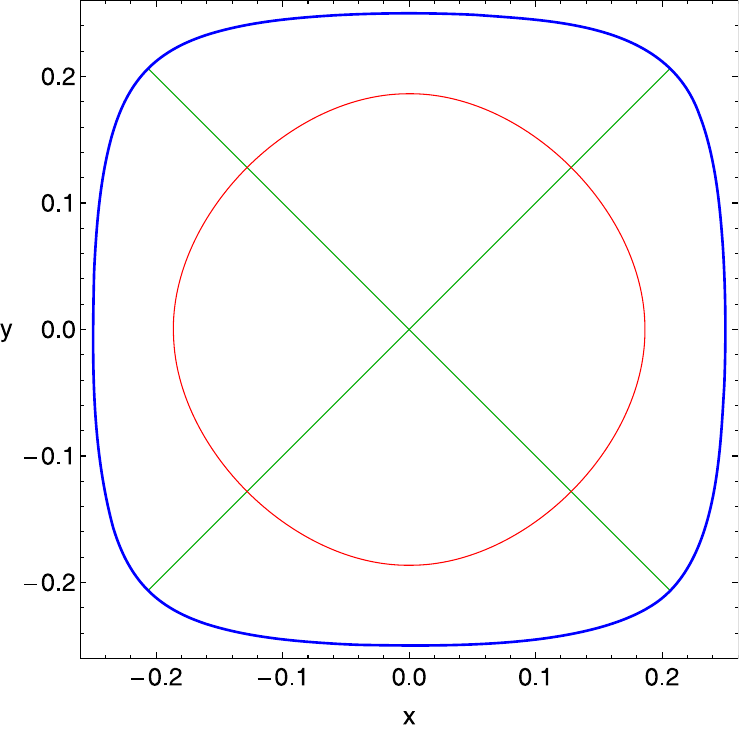}}}
\caption{The 1:1 stable and unstable periodic orbits when the value of the energy is 0.005. The initial conditions of the periodic orbits are given in the text.}
\label{orb11}
\end{figure}

The plots of the two types of 1:1 resonant periodic orbits are given in Fig. \ref{orb11}. The stable orbits are the two straight-lines with initial conditions: $x_0 = 0$, $y_0 = 0$, $\dot{x_0} = \pm 0.070710$, while the initial value of $\dot{y_0}$ is always obtained from the energy integral (\ref{ham}). On the other hand, the unstable circular periodic orbit has initial conditions: $x_0 = 0.186387$, $y_0 = 0$, $\dot{x_0} = 0$. The outermost blue curve is the limiting curve at the $(x, y)$ plane and is calculated as
\begin{equation}
f_2(x,y) = V(x,y) = h.
\label{ZVCxy}
\end{equation}

Here we must point out, that for the motion along the lines
\begin{equation}
y = k x,
\label{genlins}
\end{equation}
the equations of motion (\ref{eqmot}) become identical, since both frequencies of the oscillations are always equal. Thus, the straight lines (\ref{genlins}) are also exact 1:1 periodic orbits going through the origin.

\begin{figure}[!tH]
\centering
\resizebox{\hsize}{!}{\rotatebox{0}{\includegraphics*{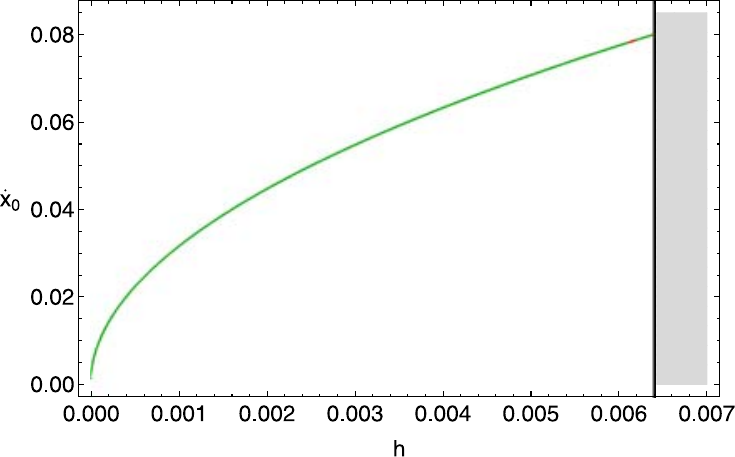}}}
\caption{Evolution of the starting position $\dot{x_0}$ of the 1:1 straight-line periodic orbits as a function of the energy $h$. Throughout the paper, green dots always correspond to stable periodic orbits, while red dots indicate unstable periodic orbits.}
\label{res11Sevol}
\end{figure}

\begin{figure}[!tH]
\centering
\resizebox{\hsize}{!}{\rotatebox{0}{\includegraphics*{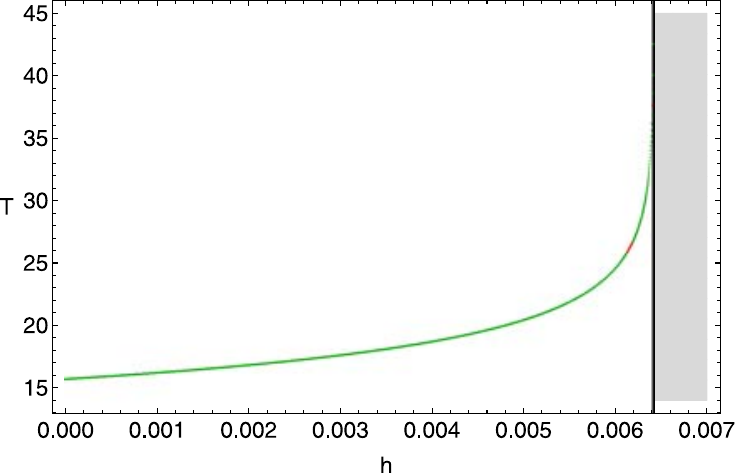}}}
\caption{Evolution of the period $T$ of the 1:1 straight-line periodic orbits as a function of the energy $h$.}
\label{T11Sevol}
\end{figure}

The evolution of the family of the 1:1 straight-line periodic orbits is presented in Fig. \ref{res11Sevol}. We see that the value of the starting position $\dot{x_0}$ increases as we increase the energy. The black vertical line indicates the energy of escape which is $h_{esc} = 0.0064$. According to the diagram, the vast majority of the computed periodic orbits are stable and therefore, are represented by green dots. Throughout the paper, green dots always correspond to stable periodic orbits, while red dots indicate unstable periodic orbits. For easy reference we may call this family as F11S. The family becomes unstable (red dots) in the interval $0.00611 < h < 0.00617$ and if we use smaller energy step we will see, that in fact, there are infinite transitions to instability and stability until $h = h_{esc}$. In this case, the escape energy acts as a barrier ending the evolution of the family without having penetration in the gray shaded area. The starting position of the 1:1 straight-line periodic orbits can be computed analytically is are given by Eq. (\ref{pxs}).

The evolution of the period of the same family of periodic orbits is presented in Fig. \ref{T11Sevol}. The periodic orbits terminate at $h = h_{esc}$ when their period is infinite, in accordance with the Str\"{o}mgren termination principle [\citealp{43}]. For $h > h_{esc}$ the ZVC opens and the 1:1 straight-line periodic orbits escape immediately to infinity. Apart from the semi-numerical equation (\ref{Ts}), the period of the above straight-line periodic orbits can be found analytically using the following relation
\begin{equation}
T = \frac{4}{\alpha \sqrt{\varepsilon}} F\left(\frac{\pi}{2}, k\right),
\label{Tanal}
\end{equation}
where $F\left(\pi/2, k\right)$ is the complete elliptic integral of the first kind, $k=b/\alpha$, while $\alpha$ and $b$ are the roots of the equation $\varepsilon x^4 - \omega^2 x^2 + h = 0$ (see Paper I for more details).

\begin{table}
\centering
\caption{Position and period for five sample 1:1 straight-line periodic orbits. Subscript $s$ indicates values derived theoretically using Eqs. (\ref{pxs}) and (\ref{Tanal}), while subscript $n$ shows results obtained by the numerical integration.}
\begin{tabular}{|c||c|c|c|c|}
\hline
$h$  & $\dot{x_s}$  & $\dot{x_n}$  & $T_s$  & $T_n$ \\
\hline \hline
0.0015 & 0.038729 & 0.038729 & 16.504135 & 16.504134 \\
\hline
0.0025 & 0.050000 & 0.050000 & 17.199126 & 17.199127 \\
\hline
0.0035 & 0.059160 & 0.059160 & 18.119501 & 18.119500 \\
\hline
0.0045 & 0.067082 & 0.067082 & 19.465131 & 19.465130 \\
\hline
0.0055 & 0.074161 & 0.074161 & 21.918290 & 21.918291 \\
\hline
\end{tabular}
\label{res11S}
\end{table}

In this case, both the position and the period of the 1:1 straight-line periodic orbits can be obtained analytically. Therefore, there is no need to use fourth order polynomial fitting curves. Table \ref{res11S} gives the position and the period of the 1:1 straight line periodic orbits for five values of the energy $h$. Subscript $n$ indicates values found by the numerical integration, while subscript $s$ indicates values found using Eqs. (\ref{pxs}) and (\ref{Tanal}). One may see, that in all cases the differences in the position and also in the period of the periodic orbits are almost negligible.

\begin{figure}[!tH]
\centering
\resizebox{\hsize}{!}{\rotatebox{0}{\includegraphics*{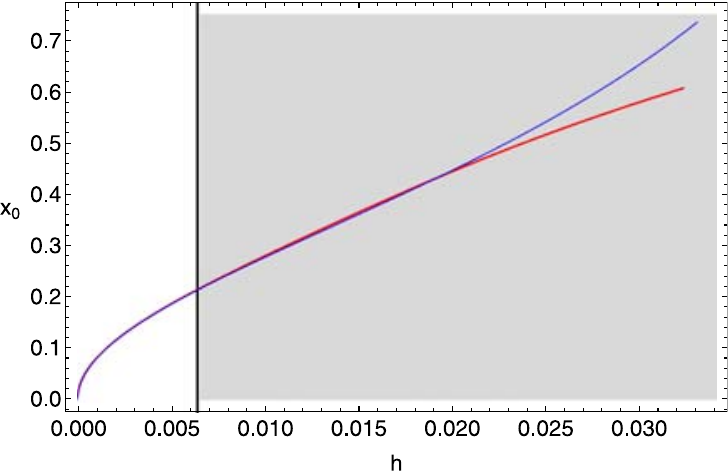}}}
\caption{Evolution of the starting position $x_0$ of the 1:1 circular periodic orbits as a function of the energy $h$.}
\label{res11Uevol}
\end{figure}

\begin{figure}[!tH]
\centering
\resizebox{\hsize}{!}{\rotatebox{0}{\includegraphics*{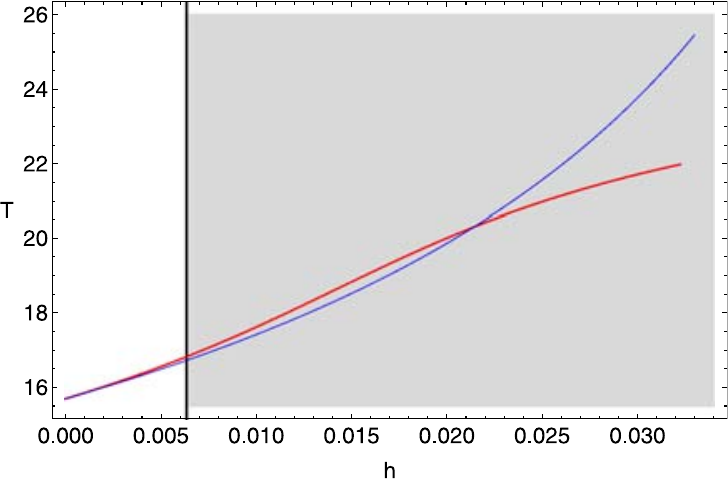}}}
\caption{Evolution of the period $T$ of the 1:1 circular periodic orbits as a function of the energy $h$.}
\label{T11Uevol}
\end{figure}

Now let's study the evolution of the unstable circular 1:1 resonant periodic orbits. Again, for easy reference we shall name this family F11U. The evolution of the F11U family is presented in Fig. \ref{res11Uevol}. One may see, that the value of the starting position $x_0$ increases with increasing energy, while all the computed periodic orbits are unstable. Moreover, we observe that the energy of escape cannot stop the evolution of the F11U family. Therefore, according to the diagram the family continues to exist for values of energy much more larger than the escape energy. The starting position $x_0$ of the 1:1 circular periodic orbits can be obtained using Eq. (\ref{xs}). Our numerical calculations suggest that in order to achieve the best agreement between theoretical and numerical outcomes, we should use the following correction term $c = 3\omega^2$. Therefore, combining Eqs. (\ref{freq}) and (\ref{xs}) and of course take into account the correction term the relation for the starting position of the F11U family becomes
\begin{equation}
x_s = \sqrt{\frac{h}{\omega^2 - 3\varepsilon h}}.
\label{xs11U}
\end{equation}
The plot of Eq. (\ref{xs11U}) is shown as the blue line in Fig. \ref{res11Uevol}. It is evident, that especially for small values of the energy the semi-numerical formula (\ref{xs11U}) can approximate the evolution of the F11U family with sufficient accuracy. On the other hand, when $h > 0.0214$ the relative error between the theoretical and the corresponding numerical results become significantly larger than 1\% and therefore, the use of the fourth order fitting curve becomes imperative.

\begin{table*}
\centering
\caption{Position and period for five sample 1:1 circular periodic orbits composing the family F11U. Subscript $n$ indicates values found by the numerical integration, subscript $s$ indicates values found using the semi-numerical relations, while subscript $f$ corresponds to values obtained using the fourth order polynomial fitting curve.}
\begin{tabular}{|c||c|c|c|c|c|c|}
\hline
$h$  & $x_s$  & $x_f$  & $x_n$  & $T_s$  & $T_f$  & $T_n$ \\
\hline \hline
0.006 & 0.205556 & 0.206582 & 0.206556 & 16.673842 & 16.750468 & 16.762218 \\
\hline
0.012 & 0.311085 & 0.316962 & 0.314721 & 17.843047 & 18.119173 & 18.100850 \\
\hline
0.018 & 0.411081 & 0.412821 & 0.414016 & 19.298650 & 19.547718 & 19.561114 \\
\hline
0.024 & 0.522232 & 0.503353 & 0.503286 & 21.180613 & 20.798522 & 20.798931 \\
\hline
0.030 & 0.654653 & 0.581423 & 0.580445 & 23.748208 & 21.707098 & 21.701496 \\
\hline
\end{tabular}
\label{res11U}
\end{table*}

\begin{figure}[!tH]
\centering
\resizebox{\hsize}{!}{\rotatebox{0}{\includegraphics*{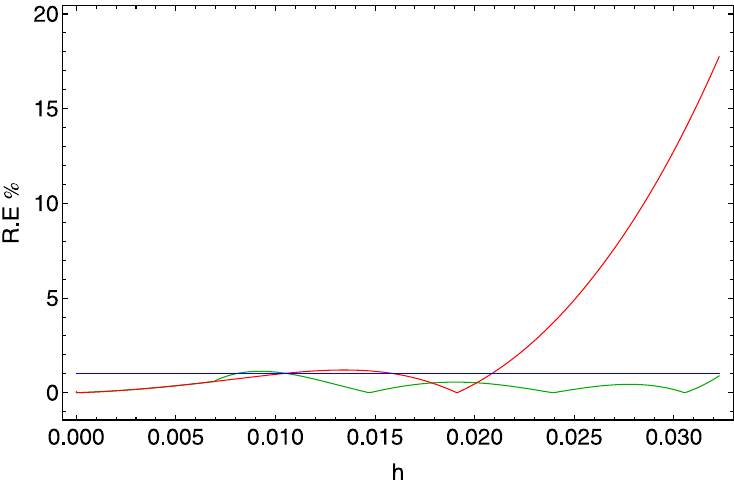}}}
\caption{The evolution of the relative error of the semi-numerical relation and the polynomial fitting curve as a function 
of the energy for the position $x_0$ of the periodic orbits composing F11U family. Red color is used for the relative error obtained by the semi-numerical formulas, while green color corresponds to the polynomial fitting curves. The horizontal blue line indicates the accuracy threshold value (1\%).}
\label{errs_x}
\end{figure}

\begin{figure}[!tH]
\centering
\resizebox{\hsize}{!}{\rotatebox{0}{\includegraphics*{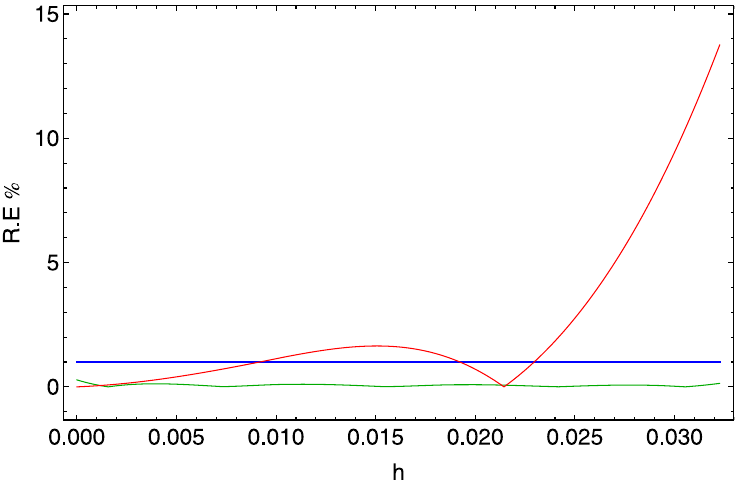}}}
\caption{The evolution of the relative error of the semi-numerical relation and the polynomial fitting curve as a function of the energy for the period $T$ of the periodic orbits composing F11U family.}
\label{errs_T}
\end{figure}

The evolution of the period of the F11U family is presented in Fig. \ref{T11Uevol}. The blue line corresponds to the semi-numerical formula (\ref{Ts}). We observe, that the agreement is sufficient only when $h < h_{esc}$. Thus, the fourth order polynomial fitting curve is the only solution in order to obtain semi-numerically the period of the orbits which belong to the F11U family. Table \ref{res11U} gives the position and the period of the 1:1 circular periodic orbits for five values of the energy $h$. Once more, subscript $n$ indicates values found by the numerical integration, subscript $s$ indicates values found using the semi-numerical relations, while subscript $f$ corresponds to values obtained using the fourth order polynomial fitting curve. We see, that in all cases the results obtained by the fitting curve are much more accurate than those derived using the semi-numerical relations. The terms of the fourth order polynomial fitting curves giving the position and the period of the orbits of the F11U family can be found in Tables (\ref{fitx}) and (\ref{fitT}) which are given at the end of this Section.

In order to prove beyond any doubt the efficiency of the fourth-order polynomial fitting curves, we present in Figs. \ref{errs_x} and Fig. \ref{errs_T} the evolution of the relative error for both the position and the period of the periodic orbits respectively, using (i) the semi-numerical formulas and (ii) the fourth-order polynomial fitting curves. With red color we plot the relative error obtained by the semi-numerical formulas, while green color corresponds to the polynomial fitting curves. The horizontal blue line indicates the accuracy threshold value (1\%). At this point, we should define what we mean by the term ``relative error". Let $x_n$ be the true value of the position of a periodic orbit, which has been obtained using the numerical integration. Our target is to approximate this value. Therefore, $x_s$ is the value derived by the semi-numerical relation and $x_f$ is the value from the polynomial fitting curve. The relative error corresponding to the semi-numerical formula is defined as $R.E = |x_n - x_s|/x_n$, while for the fitting curve is $R.E = |x_n - x_f|/x_n$. Using the same philosophy we can define similar relative errors for the period of the periodic orbits. It is evident from Fig. \ref{errs_x} that for small values of the energy, both methods can provide reliable results regarding the position of the periodic orbits. However, when $h > 0.022$ the semi-numerical formula cannot follow the evolution of the F11U family yielding to prohibiting relative error. On the other hand, the polynomial fitting curve can approximate equally well the entire family. Things are quite similar in Fig. \ref{errs_T} where we observe that the relative error of the polynomial fitting curve tends practically to zero, while the semi-numerical relation is valid only when $h < 0.024$.

\subsection{The 1:2 resonance}
\label{res12}

\begin{figure}[!tH]
\centering
\resizebox{\hsize}{!}{\rotatebox{0}{\includegraphics*{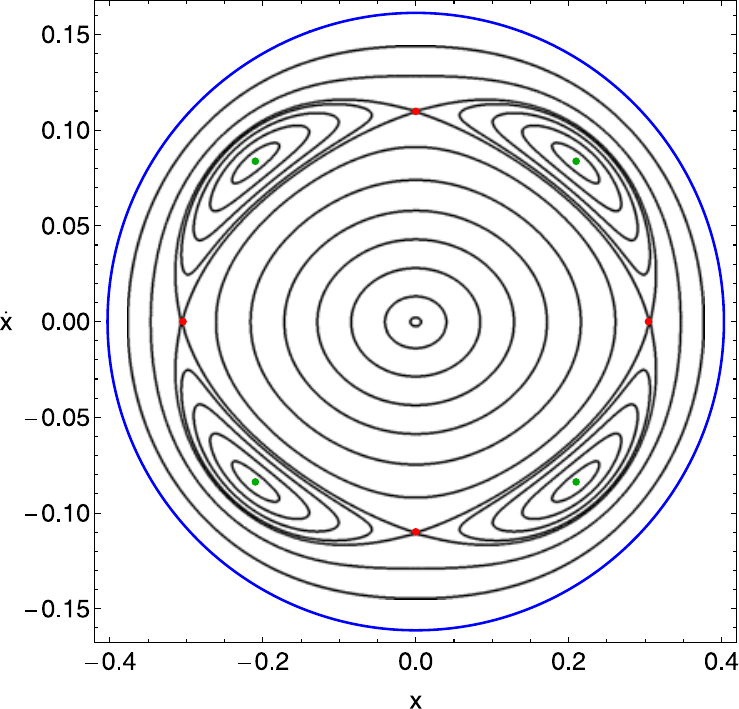}}}
\caption{The $(x, \dot{x})$ phase plane for the 1:2 resonance case. Here, $\omega_1 = 0.4$, $\omega_2 = 0.8$, while the value of the energy is 0.013.}
\label{pss12}
\end{figure}

\begin{figure*}[!tH]
\centering
\resizebox{\hsize}{!}{\rotatebox{0}{\includegraphics*{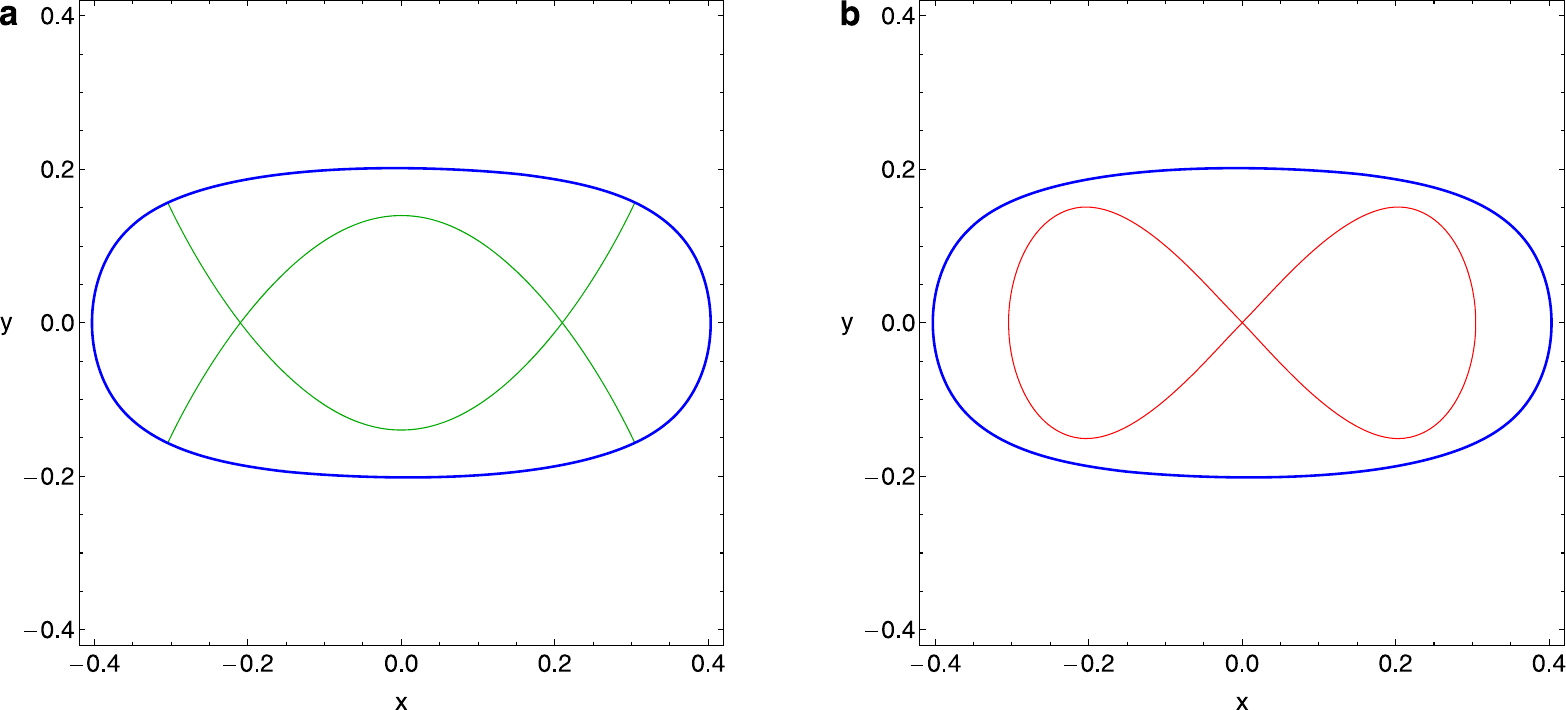}}}
\caption{(a-b): (a-left): Two symmetrical 1:2 stable periodic orbits and (b-right): an unstable 1:2 periodic orbit when the value of the energy is 0.013. The initial conditions of the periodic orbits are given in the text.}
\label{orbs12}
\end{figure*}

We now proceed to the 1:2 resonance when $\omega_1 = 0.4$ and $\omega_2 = 0.8$. Fig. \ref{pss12} shows the $(x, \dot{x})$ PSS when $h = 0.013$. We see that the phase plane is covered entirely by regular orbits. Looking at the same plot, we can identify the location of two types of 1:2 resonant periodic orbits. The first type which corresponds to the stable orbits with four symmetrical starting positions indicated by green dots. These points belong to the straight lines $\dot{x} = \pm \omega_1 x$. On the other hand, the second type has also four starting positions marked with red dots of which two of them are located at the $x$-axis and two of them at the $\dot{x}$-axis. All four points correspond to the same unstable periodic orbit which in fact produces the separatrix. Similarly following the same reasoning as in the case of the 1:1 resonance, we can prove that all the starting positions for both stable and unstable 1:2 periodic orbits belong to the ellipse $x^2/|x_0|^2 + \dot{x}^2/\dot{|x_0|}^2 = 1$, where $|x_0|$ and $\dot{|x_0|}$ are the starting positions of the unstable periodic orbit at the $x$-axis and $\dot{x}$-axis respectively. The two types of 1:2 resonant periodic orbits are presented in Fig. \ref{orbs12}(a-b). The two symmetrical stable periodic orbits shown in Fig. \ref{orbs12}a have initial conditions: $x_0 = \pm 0.209760$, $y_0 = 0$, $\dot{x_0} = 0.083754$, while the unstable figure-eight periodic orbit (Fig. \ref{orbs12}b) has initial conditions: $x_0 = 0.304369$, $y_0 = 0$, $\dot{x_0} = 0$.

\begin{figure*}[!tH]
\centering
\resizebox{\hsize}{!}{\rotatebox{0}{\includegraphics*{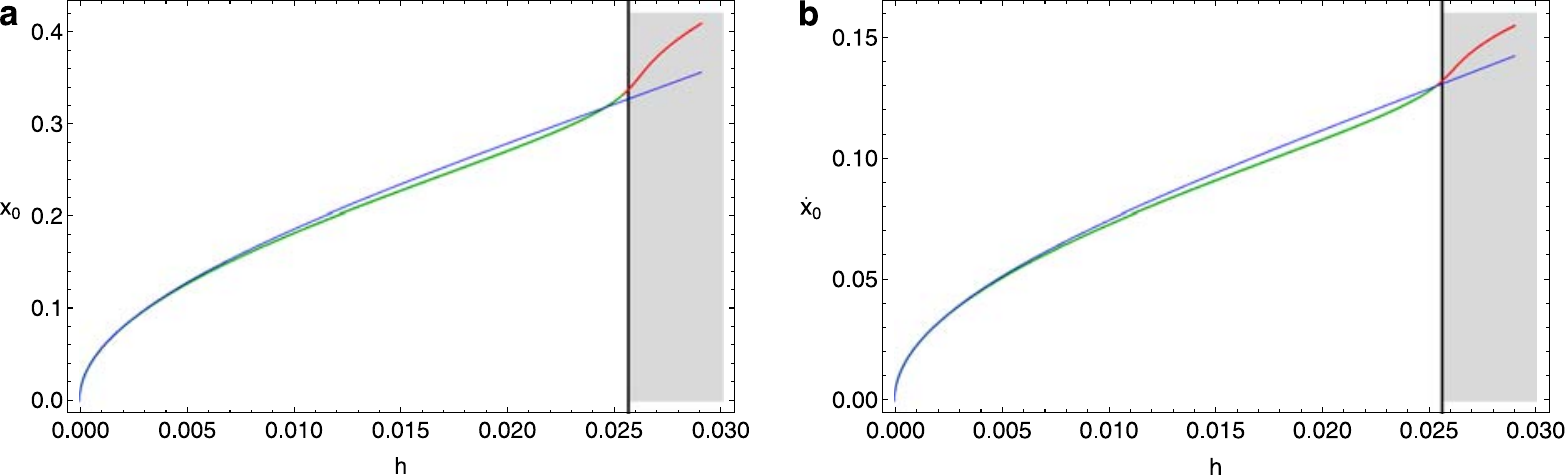}}}
\caption{(a-b): Evolution of the starting position (a-left): $x_0$ and (b-right): $\dot{x_0}$ of the periodic orbits of the F12S family as a function of the energy $h$.}
\label{res12Sevol}
\end{figure*}

\begin{figure}[!tH]
\centering
\resizebox{\hsize}{!}{\rotatebox{0}{\includegraphics*{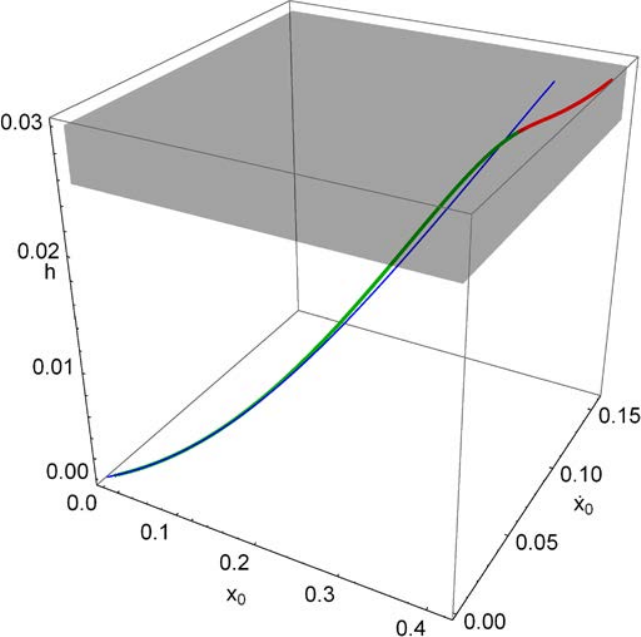}}}
\caption{Three dimensional (3D) evolution of the F12S family of periodic orbits as a function of the starting position $(x_0, \dot{x_0})$ of the periodic orbits.}
\label{res12S3d}
\end{figure}

Let us first investigate the evolution of the first type of the 1:2 resonant periodic orbits. These orbits belong to a family which we shall call F12S. Studying the evolution of this particular family of periodic orbits is indeed a real challenge due to the peculiar nature of these orbits. The orbits which consist the F12S family do not pass through the origin and also do not start perpendicularly from the $x$-axis. Therefore, we cannot use neither Eq. (\ref{xs}) nor Eq. (\ref{pxs}) in order to compute semi-numerically their location. Thus, we have to develop new semi-numerical relations. Our numerical experiments indicate, that the location of these orbits can be obtained using the following semi-numerical equations
\begin{eqnarray}
x_s &=& \omega_1 \sqrt{\frac{2h}{4\omega_1^4 - \varepsilon h}}, \nonumber \\
\dot{x_s} &=& \omega_1 x_s.
\label{res12Stheor}
\end{eqnarray}

\begin{figure}[!tH]
\centering
\resizebox{\hsize}{!}{\rotatebox{0}{\includegraphics*{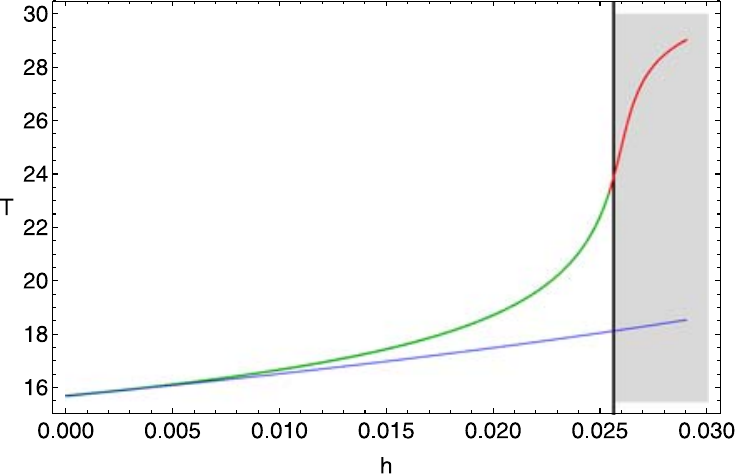}}}
\caption{Evolution of the period $T$ of the F12S family as a function of the energy $h$.}
\label{T12Sevol}
\end{figure}

\begin{figure}[!tH]
\centering
\resizebox{\hsize}{!}{\rotatebox{0}{\includegraphics*{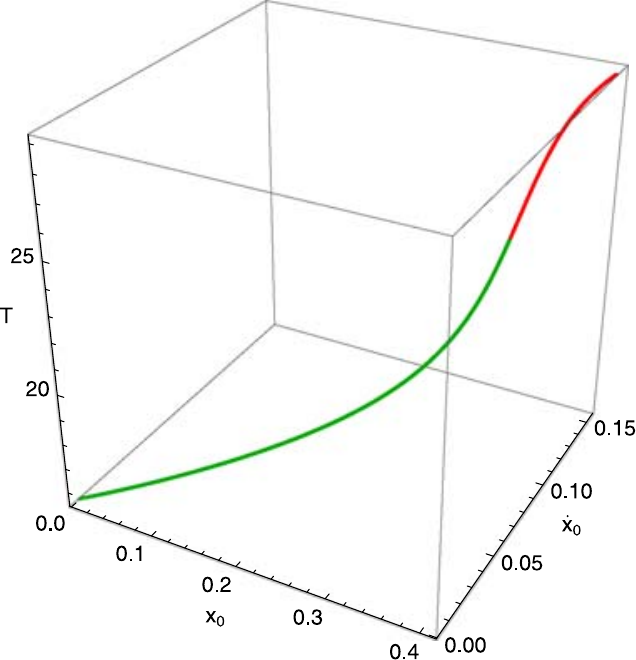}}}
\caption{Evolution of the period $T$ of the F12S family as a function of the starting position $(x_0, \dot{x_0})$ of the periodic orbits.}
\label{T12S3d}
\end{figure}

The evolution of the F12S family is presented in Fig. \ref{res12Sevol}(a-b). We see that both values of $x_0$ and $\dot{x_0}$ increase as we increase the energy. Moreover, we observe that the majority of the computed periodic orbits are stable. However, when $h > 0.02538$ that is just before reaching the escape energy $(h_{esc} = 0.0256)$ the orbits become unstable. It is evident, that in this case the escape energy cannot stop the evolution of the family which penetrates this imaginary barrier and ends inside the shaded area when $h = 0.02901$. Our semi-numerical Eqs. (\ref{res12Stheor}) are valid only when $h < h_{esc}$. In order to achieve much better agreement we should use the fourth order polynomial fitting curves. The terms of the fitting curves of the polynomials $x_f(h)$ and $\dot{x_f}(h)$ are given in Tables (\ref{fitx}) and (\ref{fitpx}) respectively. It would be very interesting to combine both Eqs. (\ref{res12Stheor}) in parametric form in order to visualize in three dimensions the evolution of the F12S family. Such a plot is presented in Fig. \ref{res12S3d}.

The evolution of the period of the F12S family is shown in Fig. \ref{T12Sevol}. The blue line corresponds to the semi-numerical formula (\ref{Ts}). In this case, the agreement is sufficient only when $h < 0.0082$. Therefore, the use  of the fourth order polynomial fitting curve becomes imperative in order to obtain semi-numerically the period of the orbits which belong to the F12S family. Fig. \ref{T12S3d} depicts an alternative approach to the evolution of the period of the F12S family. Here, the period is not given as a function of the energy $h$ but as a function of the starting position of the orbits using the initial coordinates $x_0$ and velocities $\dot{x_0}$ in a three-dimensional plot. Table \ref{res12S} gives the position and the period of the F12S family for five values of the energy $h$. The terms of the fourth order polynomial fitting curves giving the position and the period of the orbits of the F12S family are given in Tables (\ref{fitx}), (\ref{fitpx}) and (\ref{fitT}).

\begin{table*}
\centering
\caption{Position and period for five sample 1:2 periodic orbits of the F12S family.}
\begin{tabular}{|c||c|c|c|c|c|c|c|c|c|}
\hline
$h$  & $x_s$  & $x_f$  & $x_n$  & $\dot{x_s}$  & $\dot{x_f}$  & $\dot{x_n}$  & $T_s$  & $T_f$  & $T_n$ \\
\hline \hline
0.005 & 0.128168 & 0.125327 & 0.126607 & 0.051267 & 0.050270 & 0.050632 & 16.106099 & 16.101307 & 16.136088 \\
\hline
0.010 & 0.186096 & 0.184671 & 0.181873 & 0.074438 & 0.073639 & 0.072680 & 16.536129 & 16.782913 & 16.687941 \\
\hline
0.015 & 0.234350 & 0.226234 & 0.227339 & 0.093739 & 0.090236 & 0.090699 & 17.002553 & 17.359103 & 17.461039 \\
\hline
0.020 & 0.278693 & 0.268011 & 0.270559 & 0.111477 & 0.106803 & 0.107558 & 17.510810 & 18.671458 & 18.740142 \\
\hline
0.025 & 0.331238 & 0.321495 & 0.324827 & 0.129675 & 0.128598 & 0.129675 & 18.067554 & 22.952936 & 22.483271 \\
\hline
\end{tabular}
\label{res12S}
\end{table*}

\begin{figure}[!tH]
\centering
\resizebox{\hsize}{!}{\rotatebox{0}{\includegraphics*{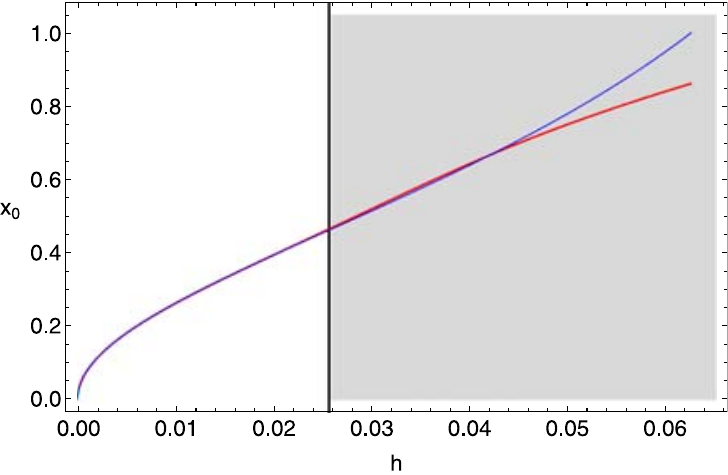}}}
\caption{Evolution of the starting position $x_0$ of the F12U family as a function of the energy $h$.}
\label{res12Uevol}
\end{figure}

\begin{figure}[!tH]
\centering
\resizebox{\hsize}{!}{\rotatebox{0}{\includegraphics*{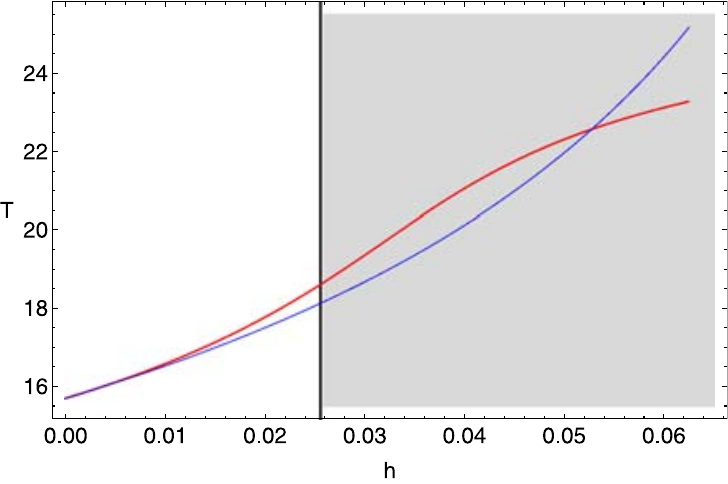}}}
\caption{Evolution of the period $T$ of the F12U family as a function of the energy $h$.}
\label{T12Uevol}
\end{figure}

The following case under investigation, is the evolution of the unstable 1:2 resonant periodic orbits. These periodic orbits form a family which we shall name F12U. The evolution of the F12U family is presented in Fig. \ref{res12Uevol}. We see, that the value of the starting position $x_0$ increases with increasing energy, while the vast majority of the computed periodic orbits are unstable. However, for extremely small values of the energy $(h < 0.000129)$ that is when we are close to the lower limit of the family, the orbits become stable. Even if family F12U consists mainly of unstable periodic orbits, the energy of escape is not able to intercept the evolution of the family, which penetrates easily the $h_{esc}$ barrier and ends only when $h \simeq 2.42 h_{esc}$.

\begin{table*}
\centering
\caption{Position and period for five sample 1:2 periodic orbits of the F12U family.}
\begin{tabular}{|c||c|c|c|c|c|c|}
\hline
$h$  & $x_s$  & $x_f$  & $x_n$  & $T_s$  & $T_f$  & $T_n$ \\
\hline \hline
0.01 & 0.259772 & 0.263180 & 0.262482 & 16.536129 & 16.543654 & 16.589184 \\
\hline
0.02 & 0.399125 & 0.394131 & 0.394614 & 17.510810 & 17.813812 & 17.776954 \\
\hline
0.03 & 0.514969 & 0.520237 & 0.520434 & 18.681019 & 19.382013 & 19.358636 \\
\hline
0.04 & 0.639831 & 0.640512 & 0.644058 & 20.122297 & 20.986128 & 21.046991 \\
\hline
0.05 & 0.781464 & 0.754361 & 0.751658 & 21.958577 & 22.331711 & 22.300594 \\
\hline
\end{tabular}
\label{res12U}
\end{table*}

The starting position $x_0$ of the periodic orbits of this family can be obtained using Eq. (\ref{xs}). Our numerical calculations suggest, that in order to achieve the best agreement between theoretical and numerical outcomes, the correction term should be $c = 1$. Thus, combining Eqs. (\ref{freq}) and (\ref{xs}) and taking into account that $\omega_2 = 2 \omega_1$ and $m = 2n$ the relation for the starting position of periodic orbits of the F12U family becomes
\begin{equation}
x_s = 2\omega_1 \sqrt{\frac{h}{4\omega_1^4 - \varepsilon h}}.
\label{xs12U}
\end{equation}
The plot of Eq. (\ref{xs12U}) is shown as the blue line in Fig. \ref{res12Uevol}. The semi-numerical formula (\ref{xs12U}) can approximate the evolution of the F12U family with sufficient accuracy only when $h < 0.0442$. For larger values of the energy the relative error between the theoretical and the corresponding numerical results become significantly larger than 1\% and therefore, the use of the fourth order fitting curve becomes imperative.

The evolution of the period of the F12U family is presented in Fig. \ref{T12Uevol}. The blue line corresponds to the semi-numerical formula (\ref{Ts}). The agreement is sufficient only when $h < 0.0124$. Thus, the fourth order polynomial fitting curve is the only solution in order to obtain semi-numerically the period of the orbits which belong to the F12U family. Table \ref{res12U} gives the position and the period of the F12U family for five values of the energy $h$.

\subsection{The 1:3 resonance}
\label{res13}

\begin{figure}[!tH]
\centering
\resizebox{\hsize}{!}{\rotatebox{0}{\includegraphics*{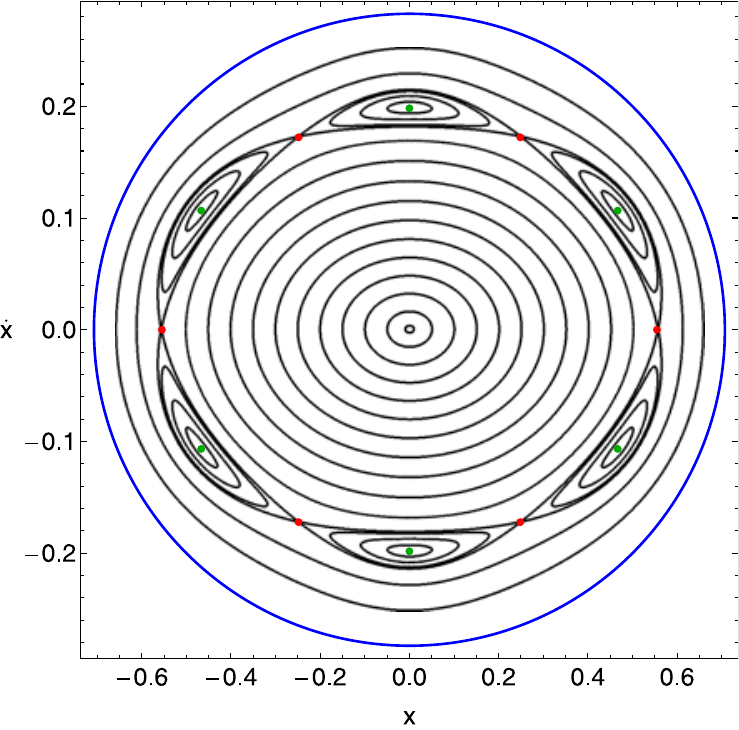}}}
\caption{The $(x, \dot{x})$ phase plane for the 1:3 resonance case. Here, $\omega_1 = 0.4$, $\omega_2 = 1.2$, while the value of the energy is 0.04.}
\label{pss13}
\end{figure}

\begin{figure*}[!tH]
\centering
\resizebox{\hsize}{!}{\rotatebox{0}{\includegraphics*{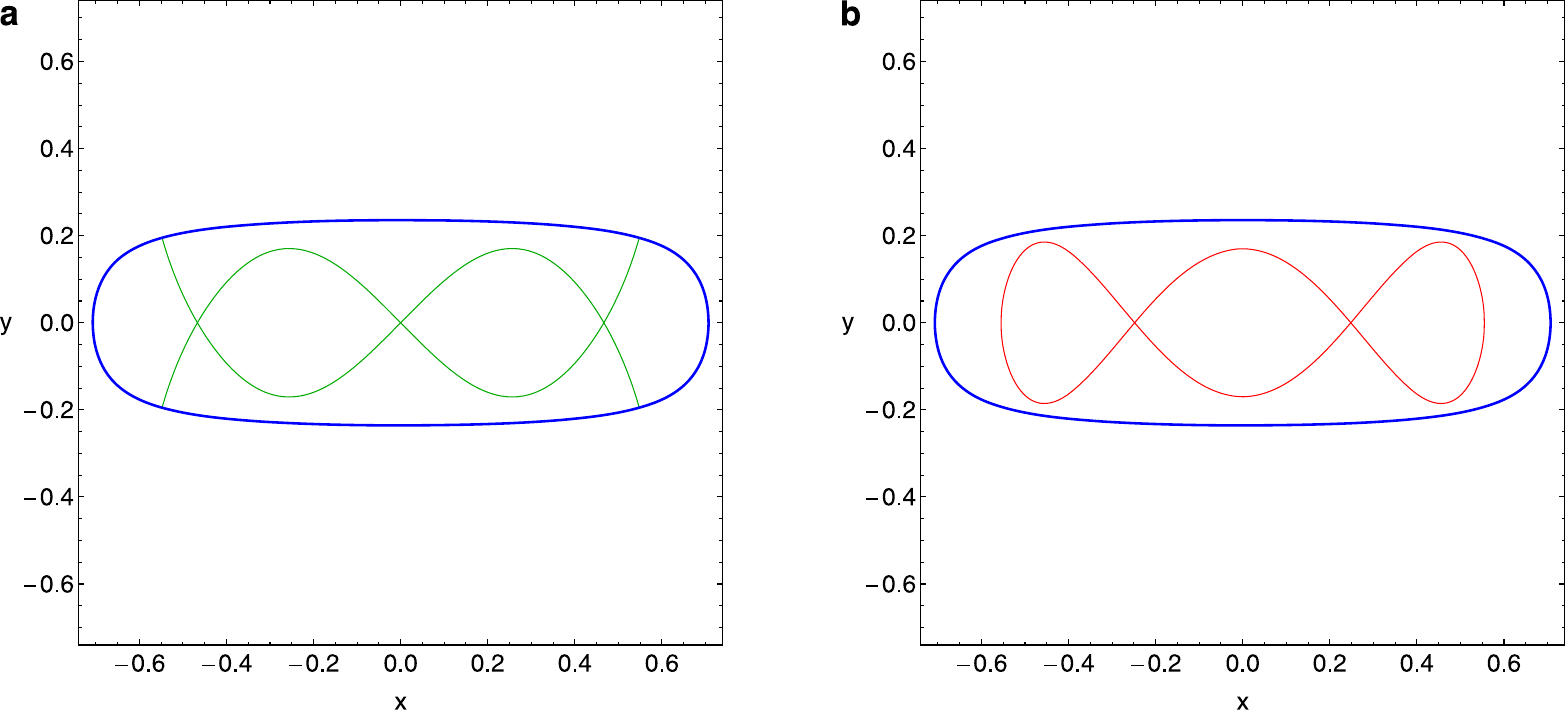}}}
\caption{(a-b): (a-left): Two symmetrical 1:3 stable periodic orbits and (b-right): an unstable 1:3 periodic orbit when the value of the energy is 0.04. The initial conditions of the periodic orbits are given in the text.}
\label{orbs13}
\end{figure*}

Another interesting resonance case is the 1:3 resonance when $\omega_1 = 0.4$ and $\omega_2 = 1.2$. Fig. \ref{pss13} shows the $(x, \dot{x})$ phase plane when the value of the energy is $h = 0.04$. We observe, that the phase plane is covered entirely by regular orbits, while two types of 1:3 resonant periodic orbits can be identified. The first type which corresponds to the stable orbits has six symmetrical starting positions indicated by green dots. On the other hand, the second type has also six starting positions marked with red dots and correspond to the same unstable periodic orbit which in fact produce the separatrix. As in the previous cases, all the starting positions for both stable and unstable 1:3 periodic orbits belong to the ellipse $x^2/|x_0|^2 + \dot{x}^2/\dot{|x_0|}^2 = 1$, where $|x_0|$ and $\dot{|x_0|}$ are the starting positions of the unstable and stable periodic orbit respectively. The two types of 1:3 resonant periodic orbits are presented in Fig. \ref{orbs13}(a-b). The two symmetrical stable periodic orbits shown in Fig. \ref{orbs13}a have initial conditions: $x_0 = 0$, $y_0 = 0$, $\dot{x_0} = \pm 0.198262$, while the unstable periodic orbit (Fig. \ref{orbs13}b) has initial conditions: $x_0 = 0.554740$, $y_0 = 0$, $\dot{x_0} = 0$.

\begin{figure}[!tH]
\centering
\resizebox{\hsize}{!}{\rotatebox{0}{\includegraphics*{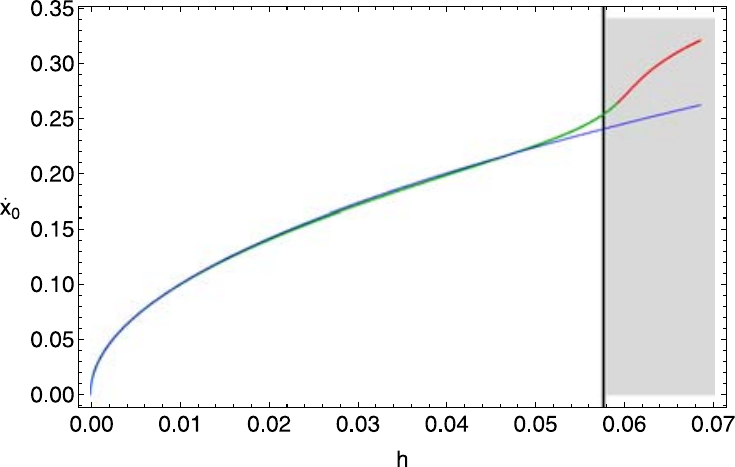}}}
\caption{Evolution of the starting position $\dot{x_0}$ of the F13S family as a function of the energy $h$.}
\label{res13Sevol}
\end{figure}

\begin{figure}[!tH]
\centering
\resizebox{\hsize}{!}{\rotatebox{0}{\includegraphics*{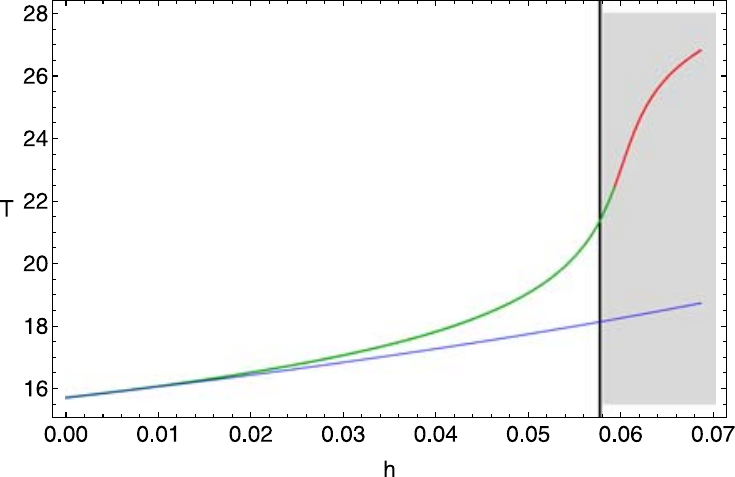}}}
\caption{Evolution of the period $T$ of the F13S family as a function of the energy $h$.}
\label{T13Sevol}
\end{figure}

We shall initially begin studying the evolution of the first type of the 1:3 resonant periodic orbits. These orbits compose the F13S family. Fig. \ref{res13Sevol} depicts the evolution of the F13S family of periodic orbits. The results are quite similar to the previous resonance cases. Again, the starting position $\dot{x_0}$ increases with increasing energy. The vast majority of the periodic orbits of the F13S family are stable. However, there is something very interesting and new in respect to the previous cases. We observe, that a small portion of the F13S family retains its stability even after crossing the barrier of the escape energy. In fact, stable 1:3 resonant periodic orbits still exist when $h_{esc} \leq h \preceq 0.05916$. Only when $h > 0.05916$ the periodic orbits become unstable. The starting position $\dot{x_0}$ of the periodic orbits of this family can be obtained using the semi-numerical Eq. (\ref{pxs}). The semi-numerical formula (\ref{pxs}) can approximate the evolution of the F13S family with sufficient accuracy only when $h < 0.0518$. For larger values of the energy the relative error between the theoretical and the corresponding numerical results become significantly larger than 1\% and therefore, the use of the fourth order fitting curve becomes imperative.

\begin{table*}
\centering
\caption{Position and period for five sample 1:3 periodic orbits of the F13S family.}
\begin{tabular}{|c||c|c|c|c|c|c|}
\hline
$h$  & $\dot{x_s}$  & $\dot{x_f}$  & $\dot{x_n}$  & $T_s$  & $T_f$  & $T_n$ \\
\hline \hline
0.02 & 0.142345 & 0.141421 & 0.140164 & 16.437593 & 16.568000 & 16.517926 \\
\hline
0.03 & 0.173205 & 0.171577 & 0.171299 & 16.842721 & 17.022626 & 17.071841 \\
\hline
0.04 & 0.200000 & 0.195965 & 0.198262 & 17.279359 & 17.713317 & 17.829818 \\
\hline
0.05 & 0.223606 & 0.225610 & 0.225000 & 17.751818 & 19.281189 & 19.079578 \\
\hline
0.06 & 0.244949 & 0.270370 & 0.270151 & 18.265279 & 22.747489 & 22.124797 \\
\hline
\end{tabular}
\label{res13S}
\end{table*}

Fig. \ref{T13Sevol} shows the evolution of the period of the F13S family. The blue line corresponds to the semi-numerical formula (\ref{Ts}). The agreement is sufficient only when $h < 0.0214$. Thus, the fourth order polynomial fitting curve is the only solution in order to obtain semi-numerically the period of the orbits which belong to the F13S family. Table \ref{res13S} gives the position and the period of the F13S family for five values of the energy $h$.

\begin{figure}[!tH]
\centering
\resizebox{\hsize}{!}{\rotatebox{0}{\includegraphics*{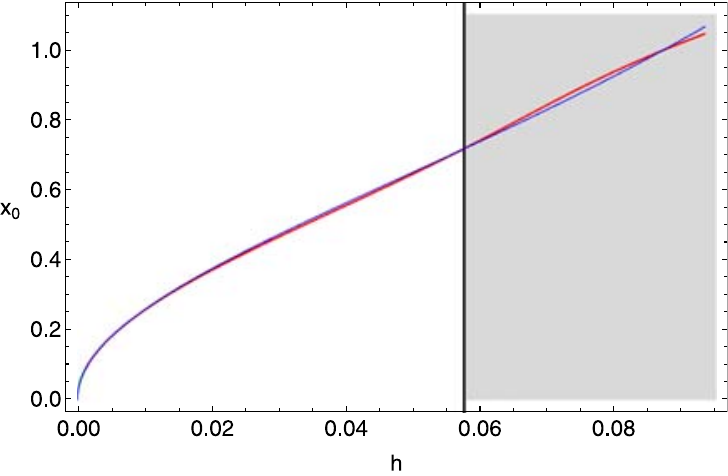}}}
\caption{Evolution of the starting position $x_0$ of the F13U family as a function of the energy $h$.}
\label{res13Uevol}
\end{figure}

\begin{figure}[!tH]
\centering
\resizebox{\hsize}{!}{\rotatebox{0}{\includegraphics*{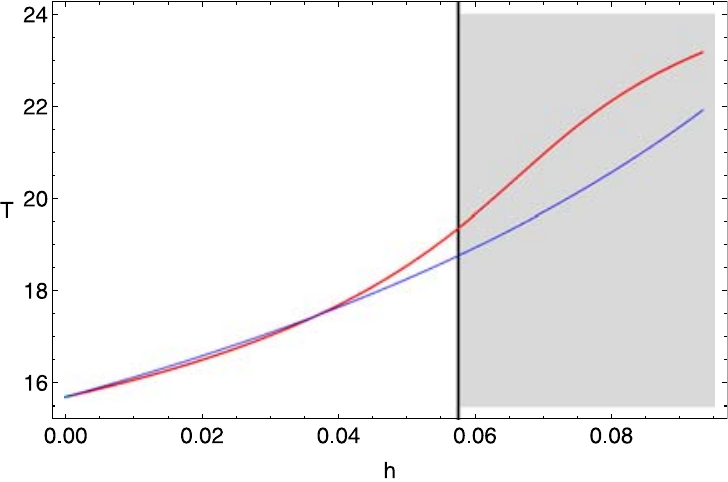}}}
\caption{Evolution of the period $T$ of the F13U family as a function of the energy $h$.}
\label{T13Uevol}
\end{figure}

We continue our research with the evolution of the unstable 1:3 resonant periodic orbits which compose the F13U family. The evolution of the F13U family is presented in Fig. \ref{res13Uevol}. It is clear, that the value of the starting position $x_0$ increases with increasing energy, while the vast majority of the computed periodic orbits are unstable. However, one may observe that for small values of the energy $(h < 0.00143)$ that is when we are very close to the lower bound of the family, the orbits become stable. Once more, the energy of escape is not able to stop the evolution of the family, which crosses the $h_{esc}$ barrier and ends only when $h = 0.09336$.

\begin{table*}
\centering
\caption{Position and period for five sample 1:3 periodic orbits of the F13U family.}
\begin{tabular}{|c||c|c|c|c|c|c|}
\hline
$h$  & $x_s$  & $x_f$  & $x_n$  & $T_s$  & $T_f$  & $T_n$ \\
\hline \hline
0.01 & 0.249384 & 0.256776 & 0.255191 & 16.133730 & 16.117981 & 16.076278 \\
\hline
0.03 & 0.471404 & 0.467897 & 0.465036 & 17.100664 & 17.000630 & 17.039220 \\
\hline
0.05 & 0.650027 & 0.643594 & 0.645265 & 18.265279 & 18.581501 & 18.549802 \\
\hline
0.07 & 0.829773 & 0.842302 & 0.842820 & 19.705641 & 20.924034 & 20.954088 \\
\hline
0.09 & 1.028992 & 1.019462 & 1.020499 & 21.551147 & 22.973320 & 22.952896 \\
\hline
\end{tabular}
\label{res13U}
\end{table*}

Since the periodic orbits of the F13U family start perpendicularly from the $x$-axis, the starting position $x_0$ can be obtained using Eq. (\ref{xs}). Our numerical calculations suggest, that in order to achieve the best agreement between theoretical and numerical outcomes, the correction term should be $c = \omega_2$. Thus, combining Eqs. (\ref{freq}) and (\ref{xs}) and taking into account that $\omega_2 = 3 \omega_1$ and $m = 3n$ the relation for the starting position of periodic orbits of the F13U family becomes
\begin{equation}
x_s = \sqrt{\frac{3\omega_1 h}{3\omega_1^3 - \varepsilon h}}.
\label{xs13U}
\end{equation}
The plot of Eq. (\ref{xs13U}) is shown as the blue line in Fig. \ref{res13Uevol}. In this case, the semi-numerical formula (\ref{xs13U}) can approximate with sufficient accuracy the evolution of the entire F13U family of periodic orbits.

Fig. \ref{T13Uevol} presents the evolution of the period of the F12U family. The blue line corresponds to the semi-numerical formula (\ref{Ts}). The agreement is sufficient only when $h < 0.0448$. Thus, the fourth order polynomial fitting curve is the only solution in order to obtain semi-numerically the period of the orbits which belong to the F13U family. Table \ref{res13U} gives the position and the period of the F13U family for five values of the energy $h$.

\subsection{The 2:3 resonance}
\label{res23}

\begin{figure}[!tH]
\centering
\resizebox{\hsize}{!}{\rotatebox{0}{\includegraphics*{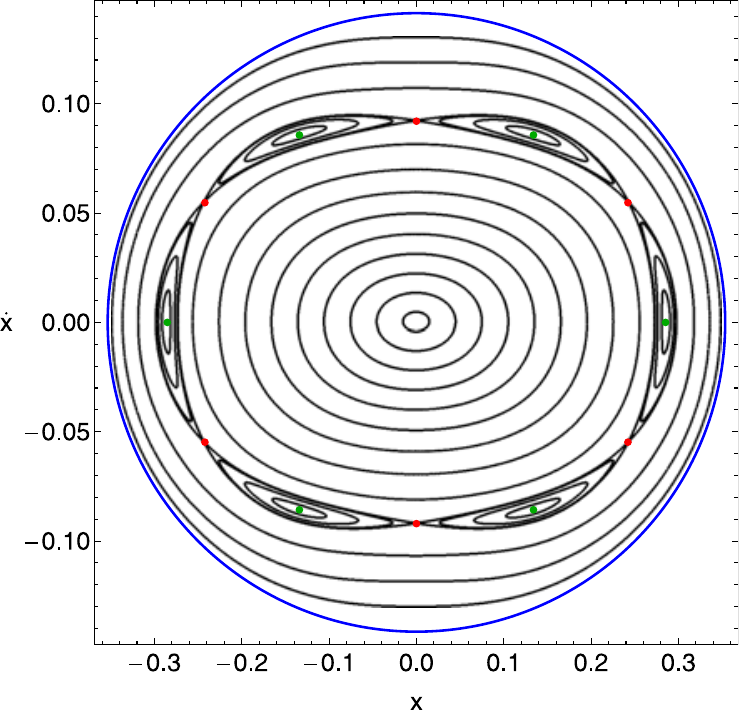}}}
\caption{The $(x, \dot{x})$ phase plane for the 2:3 resonance case. Here, $\omega_1 = 0.4$, $\omega_2 = 0.6$, while the value of the energy is 0.01.}
\label{pss23}
\end{figure}

\begin{figure*}[!tH]
\centering
\resizebox{\hsize}{!}{\rotatebox{0}{\includegraphics*{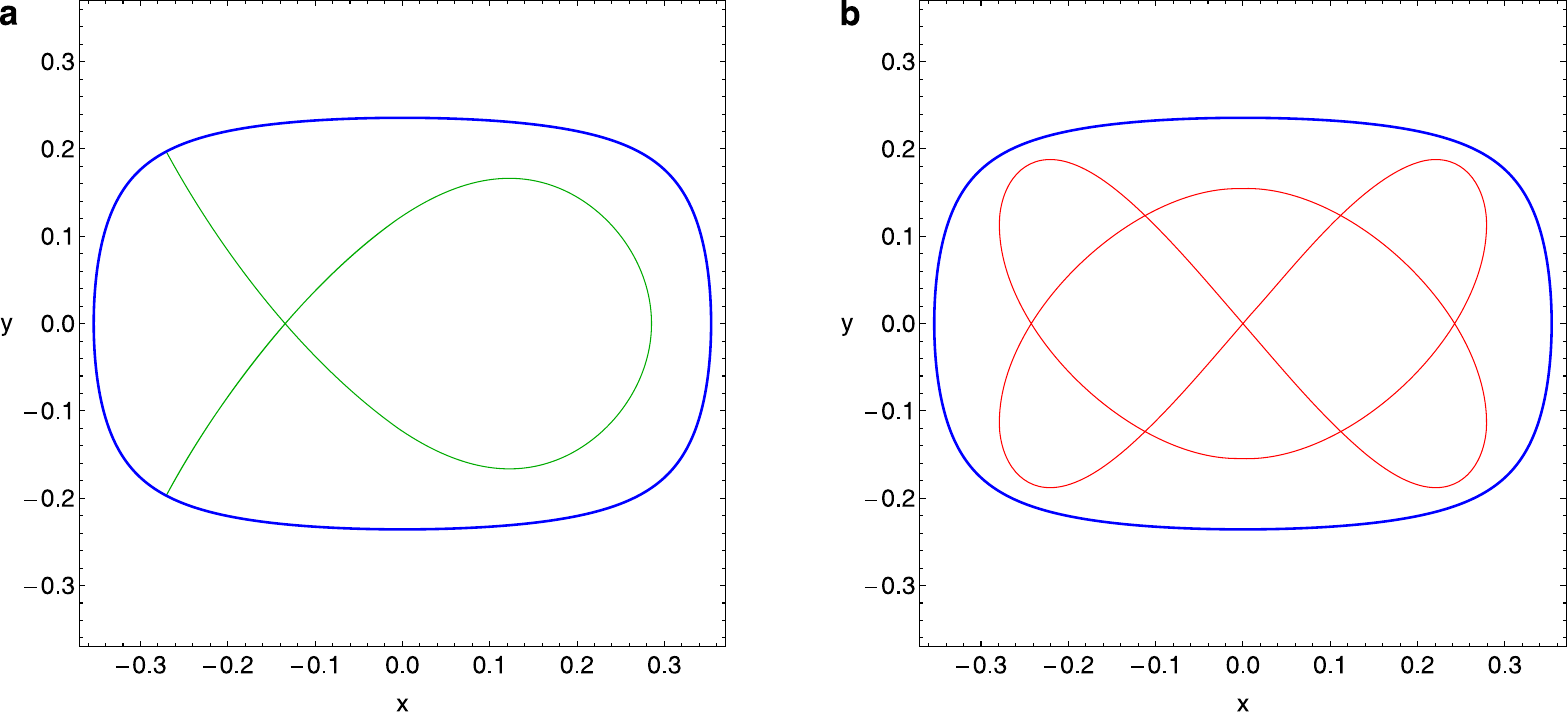}}}
\caption{(a-b): (a-left): A stable 2:3 resonant periodic orbit and (b-right): an unstable 2:3 periodic orbit when the value of the energy is 0.01. The initial conditions of the periodic orbits are given in the text.}
\label{orbs23}
\end{figure*}

We continue our investigation with the 2:3 resonance when $\omega_1 = 0.4$ and $\omega_2 = 0.6$. The $(x, \dot{x})$ PSS when the value of the energy is $h = 0.01$ is shown in Fig. \ref{pss23}. We see, that the phase plane is occupied entirely by initial conditions correspond to regular orbits. Two different types of 2:3 resonant periodic orbits can be identified. The first type which corresponds to the stable orbits has six symmetrical starting positions indicated by green dots. On the other hand, the second type has also six starting positions marked with red dots and correspond to the same unstable periodic orbit which in fact produces the separatrix. Once more, it is evident  that all the starting positions for both stable and unstable 2:3 periodic orbits belong to the ellipse $x^2/|x_0|^2 + \dot{x}^2/\dot{|x_0|}^2 = 1$, where $|x_0|$ and $\dot{|x_0|}$ are the starting positions of the stable and unstable periodic orbit respectively. The two different types of 2:3 resonant periodic orbits are presented in Fig. \ref{orbs23}(a-b). In Fig. \ref{orbs23}a we present a stable periodic orbit with initial conditions: $x_0 = 0.285273$, $y_0 = 0$, $\dot{x_0} = 0$, while the unstable periodic orbit shown in Fig. \ref{orbs23}b has initial conditions: $x_0 = 0$, $y_0 = 0$, $\dot{x_0} = 0.092018$.

\begin{figure}[!tH]
\centering
\resizebox{\hsize}{!}{\rotatebox{0}{\includegraphics*{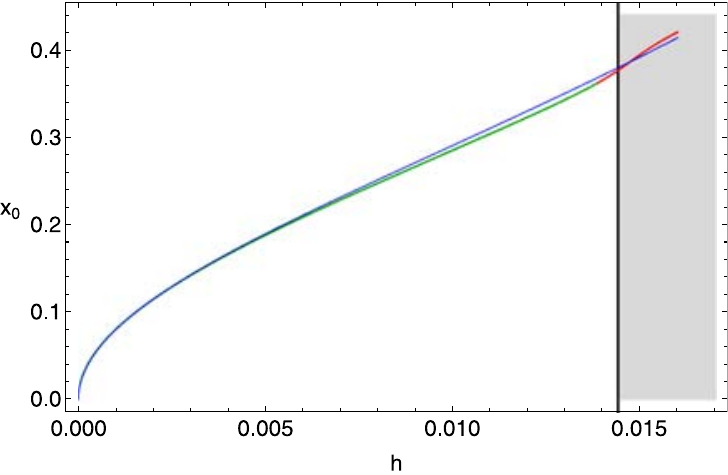}}}
\caption{Evolution of the starting position $x_0$ of the F23S family as a function of the energy $h$.}
\label{res23Sevol}
\end{figure}

\begin{figure}[!tH]
\centering
\resizebox{\hsize}{!}{\rotatebox{0}{\includegraphics*{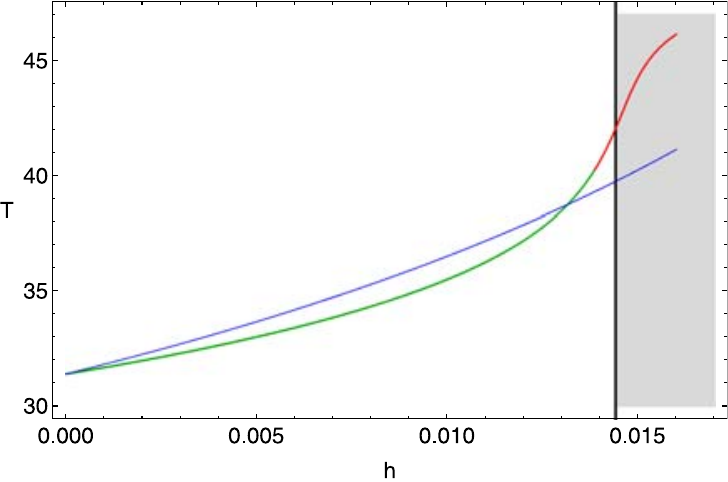}}}
\caption{Evolution of the period $T$ of the F23S family as a function of the energy $h$.}
\label{T23Sevol}
\end{figure}

Let us first shed some light on the evolution of the first type of the 2:3 resonant periodic orbits. These orbits compose the F23S family. The evolution of the F23S family of periodic orbits is presented in Fig. \ref{res23Sevol}. We see, that the starting position $x_0$ increases with increasing energy. The vast majority of the periodic orbits of the F23S family are stable. However, when $h > 0.01386$ that is just before reaching the escape energy $(h_{esc} = 0.0144)$ the orbits become unstable. Once more, the escape energy cannot stop the evolution of the F23S family which ends only when $h = 0.01598$.

\begin{table*}
\centering
\caption{Position and period for five sample 2:3 periodic orbits of the F23S family.}
\begin{tabular}{|c||c|c|c|c|c|c|}
\hline
$h$  & $x_s$  & $x_f$  & $x_n$  & $T_s$  & $T_f$  & $T_n$ \\
\hline \hline
0.003 & 0.142614 & 0.141816 & 0.141888 & 32.720042 & 32.277877 & 32.314966 \\
\hline
0.006 & 0.210818 & 0.210280 & 0.208566 & 34.201328 & 33.506128 & 33.426655 \\
\hline
0.009 & 0.271052 & 0.264906 & 0.266524 & 35.903916 & 34.802661 & 34.896645 \\
\hline
0.012 & 0.330289 & 0.323132 & 0.323187 & 37.889033 & 37.276209 & 37.208079 \\
\hline
0.015 & 0.392232 & 0.393489 & 0.394902 & 40.244594 & 43.611242 & 44.952772 \\
\hline
\end{tabular}
\label{res23S}
\end{table*}

Taking into account that the periodic orbits of the F23S family start perpendicularly from the $x$-axis, the starting position $x_0$ can be obtained using Eq. (\ref{xs}). Our numerical calculations indicate, that in order to achieve the best agreement between theoretical and numerical outcomes, the correction term should be $c = m/n$. Thus, combining Eqs. (\ref{freq}) and (\ref{xs}) and also taking into account that $\omega_2 = 3 \omega_1/2$ and $m = 3n/2$ the relation for the starting position of periodic orbits of the F23S family becomes
\begin{equation}
x_s = \omega_1 \sqrt{\frac{3h}{3\omega_1^4 - 2\varepsilon h}}.
\label{xs23S}
\end{equation}
The plot of Eq. (\ref{xs23S}) is shown as the blue line in Fig. \ref{res23Sevol}. In this case, the semi-numerical formula (\ref{xs23S}) can approximate with sufficient accuracy the evolution of the entire F23S family of periodic orbits.

\begin{figure}[!tH]
\centering
\resizebox{\hsize}{!}{\rotatebox{0}{\includegraphics*{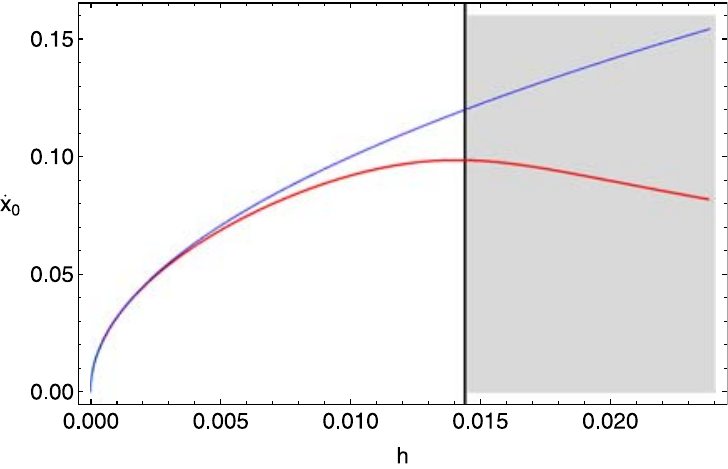}}}
\caption{Evolution of the starting position $\dot{x_0}$ of the F23U family as a function of the energy $h$.}
\label{res23Uevol}
\end{figure}

\begin{figure}[!tH]
\centering
\resizebox{\hsize}{!}{\rotatebox{0}{\includegraphics*{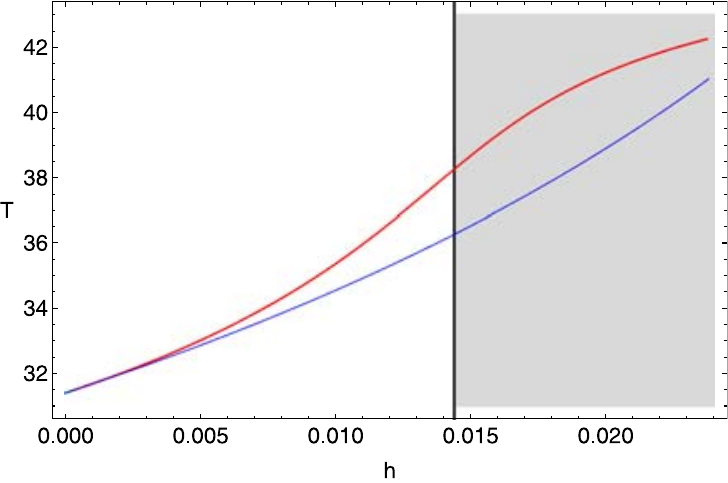}}}
\caption{Evolution of the period $T$ of the F23U family as a function of the energy $h$.}
\label{T23Uevol}
\end{figure}

In Fig. \ref{T23Sevol} we present the evolution of the period of the F23S family. The blue line corresponds to the semi-numerical formula (\ref{Ts}). We observe, that in this case the semi-numerical formula cannot be used at all, since it fails to approximate almost the entire family and the relative error between the theoretical and the corresponding numerical results become significantly larger than 1\%. Therefore, the fourth order polynomial fitting curve is the only viable solution in order to obtain semi-numerically the period of the orbits which belong to the F23S family. Table \ref{res23S} gives the position and the period of the F23S family for five values of the energy $h$.

\begin{table*}
\centering
\caption{Position and period for five sample 2:3 periodic orbits of the F23U family.}
\begin{tabular}{|c||c|c|c|c|c|c|}
\hline
$h$  & $\dot{x_s}$  & $\dot{x_f}$  & $\dot{x_n}$  & $T_s$  & $T_f$  & $T_n$ \\
\hline \hline
0.004 & 0.063245 & 0.061395 & 0.061934 & 32.567072 & 32.608581 & 32.656108 \\
\hline
0.008 & 0.089442 & 0.085900 & 0.084607 & 33.854801 & 34.342618 & 34.311747 \\
\hline
0.012 & 0.109544 & 0.096163 & 0.096830 & 35.308453 & 36.653208 & 36.605279 \\
\hline
0.016 & 0.126491 & 0.096723 & 0.097305 & 36.967020 & 39.190429 & 39.295402 \\
\hline
0.020 & 0.141421 & 0.090663 & 0.089768 & 38.883664 & 41.292766 & 41.204662 \\
\hline
\end{tabular}
\label{res23U}
\end{table*}

Our investigation proceeds with the evolution of the unstable 2:3 resonant periodic orbits which constitute the F23U family. The evolution of the F23U family is presented in Fig. \ref{res23Uevol}. We observe, that the value of the starting position $\dot{x_0}$ initially increases with increasing energy, but as we approach to the escape energy it displays a downward trend. The vast majority of the computed periodic orbits are unstable, except close to the lower limit of the family $(h < 0.00045)$ where the orbits become stable. In this case the energy of escape, as usual, cannot terminate the evolution of the F23U family, but we may say that it changes the type of evolution from increasing to decreasing. This particular peculiarity regarding the evolution of the F23U family disables the semi-numerical formula (\ref{pxs}). We can still use Eq. (\ref{pxs}) for the theoretical computation of the position $\dot{x_0}$ but only for a small part of the family when $h < 0.004$.

Fig. \ref{T23Uevol} depicts the evolution of the period of the F23U family. The blue line corresponds to the semi-numerical formula (\ref{Ts}). We observe, that in this case the semi-numerical formula can be used only when $h < 0.004$. For larger values of the energy, the relative error between the theoretical and the corresponding numerical results become significantly larger than 1\%. Therefore, we may conclude, that the fourth order polynomial fitting curves are absolutely necessary in order to obtain semi-numerically not only the period but also the starting position $\dot{x_0}$ of the orbits which belong to the F23S family. In Table \ref{res23U} we provide the position and the period of the F23U family for five values of the energy $h$.

\subsection{The 3:4 resonance}
\label{res34}

\begin{figure}[!tH]
\centering
\resizebox{\hsize}{!}{\rotatebox{0}{\includegraphics*{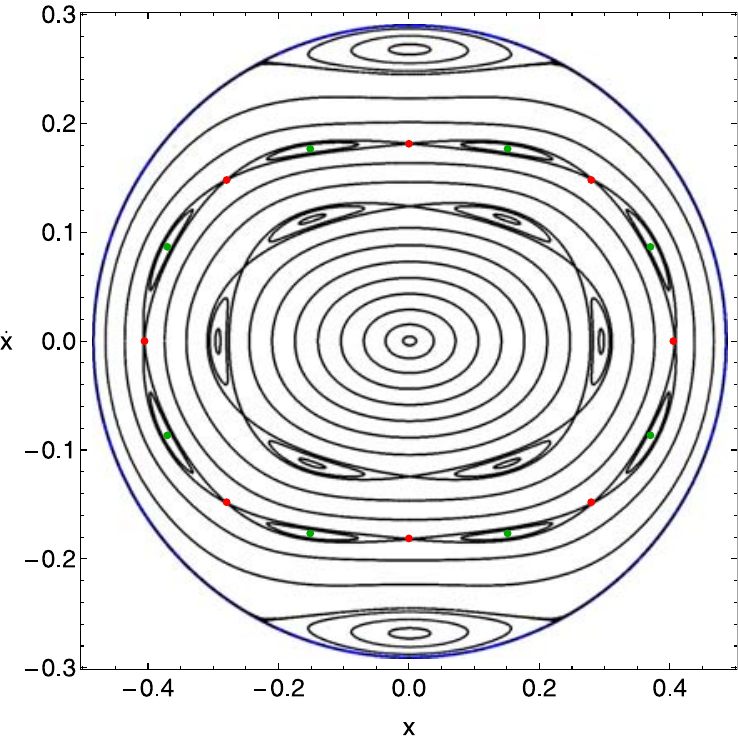}}}
\caption{The $(x, \dot{x})$ phase plane for the 3:4 resonance case. Here, $\omega_1 = 0.6$, $\omega_2 = 0.8$, while the value of the energy is 0.042.}
\label{pss34}
\end{figure}

\begin{figure*}[!tH]
\centering
\resizebox{\hsize}{!}{\rotatebox{0}{\includegraphics*{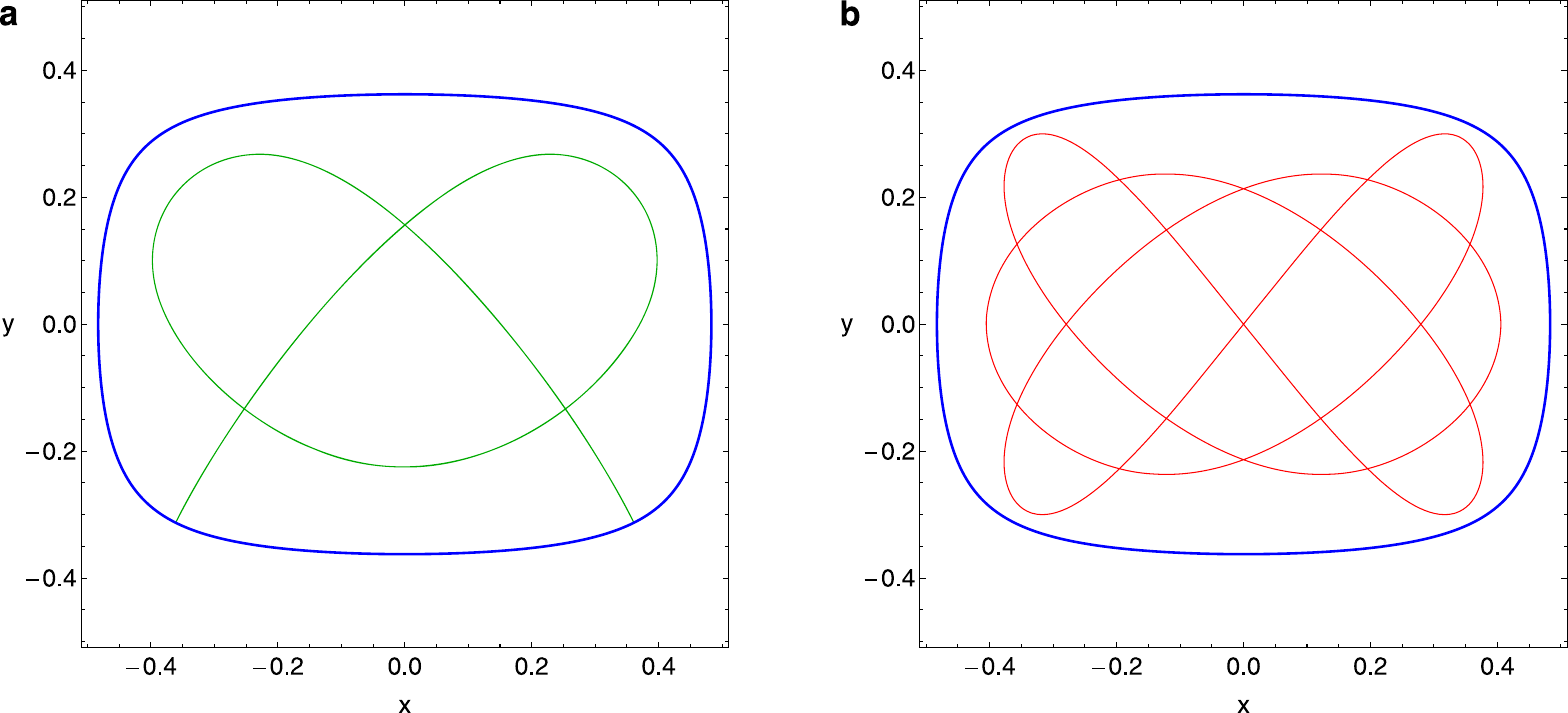}}}
\caption{(a-b): (a-left): A stable 3:4 resonant periodic orbit and (b-right): an unstable 3:4 periodic orbit when the value of the energy is 0.042. The initial conditions of the periodic orbits are given in the text.}
\label{orbs34}
\end{figure*}

The final case in our investigation is the case of the 3:4 resonance when $\omega_1 = 0.6$ and $\omega_2 = 0.8$. The $(x, \dot{x})$ phase plane when the value of the energy is $h = 0.042$ is shown in Fig. \ref{pss34}. Once more, we chose such an energy so as the phase plane to be occupied entirely by initial conditions correspond to regular orbits. Two different types of 3:4 resonant periodic orbits can be identified. The first type which corresponds to the stable orbits has eight symmetrical starting positions indicated by green dots. On the other hand, the second type has also eight starting positions marked with red dots and correspond to the same unstable periodic orbit which in fact produces the separatrix. All the starting positions for both stable and unstable 3:4 periodic orbits belong to the ellipse $x^2/|x_0|^2 + \dot{x}^2/\dot{|x_0|}^2 = 1$, where $|x_0|$ and $\dot{|x_0|}$ are the starting positions of the unstable periodic orbit on the $x$-axis and $\dot{x}$-axis respectively. Moreover we observe, that apart from the 3:4 resonance in the phase plane of Fig. \ref{pss34} the resonances 1:1 and 2:3 are also appear. The two different types of 3:4 resonant periodic orbits are presented in Fig. \ref{orbs34}(a-b). In Fig. \ref{orbs34}a we present a stable periodic orbit with initial conditions: $x_0 = 0.370222$, $y_0 = 0$, $\dot{x_0} = 0.086523$, while the unstable periodic orbit shown in Fig. \ref{orbs34}b has initial conditions: $x_0 = 0.405354$, $y_0 = 0$, $\dot{x_0} = 0$.

\begin{figure*}[!tH]
\centering
\resizebox{\hsize}{!}{\rotatebox{0}{\includegraphics*{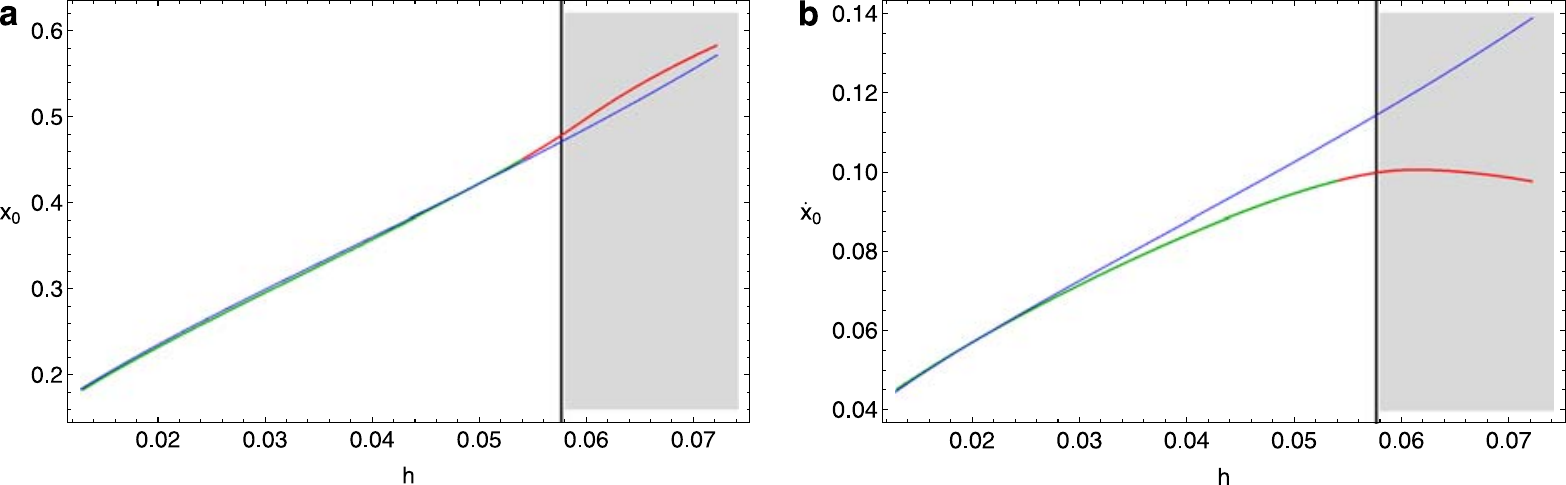}}}
\caption{(a-b): Evolution of the starting position (a-left): $x_0$ and (b-right): $\dot{x_0}$ of the periodic orbits of the F34S family as a function of the energy $h$.}
\label{res34Sevol}
\end{figure*}

\begin{figure}[!tH]
\centering
\resizebox{\hsize}{!}{\rotatebox{0}{\includegraphics*{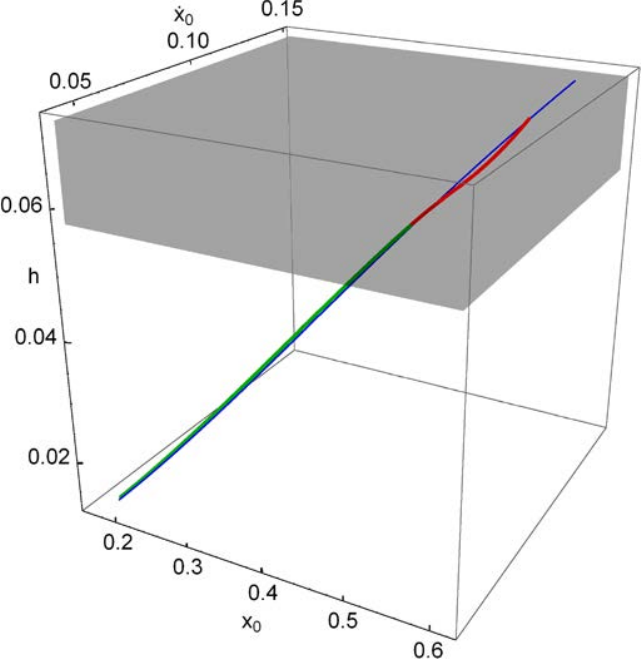}}}
\caption{Three dimensional (3D) evolution of the F34S family as a function of the starting position $(x_0, \dot{x_0})$ of the periodic orbits.}
\label{res34S3d}
\end{figure}

Let us begin our investigation with the evolution of the first type of the 3:4 resonant periodic orbits. These orbits compose a family which we shall call F34S. The periodic orbits which belong the F34S family do not pass through the origin and also do not start perpendicularly from the $x$-axis. Therefore, we cannot use neither Eq. (\ref{xs}) nor Eq. (\ref{pxs}) in order to compute semi-numerically their location. Thus, as we have seen in the case of the F12S family, we have to develop new semi-numerical relations. Our numerical experiments suggest, that the location of these orbits can be obtained using the following semi-numerical equations
\begin{eqnarray}
x_s &=& \frac{2m^2\omega_1^2}{n} \sqrt{\frac{2h}{n\left(2m^2\omega_1^4 - n^3\varepsilon h\right)}}, \nonumber \\
\dot{x_s} &=& \left(\frac{m}{n}\right)^2\omega_1^4 x_s,
\label{res34Stheor}
\end{eqnarray}
where of course, $n = 3$ and $m = 4$.

\begin{figure}[!tH]
\centering
\resizebox{\hsize}{!}{\rotatebox{0}{\includegraphics*{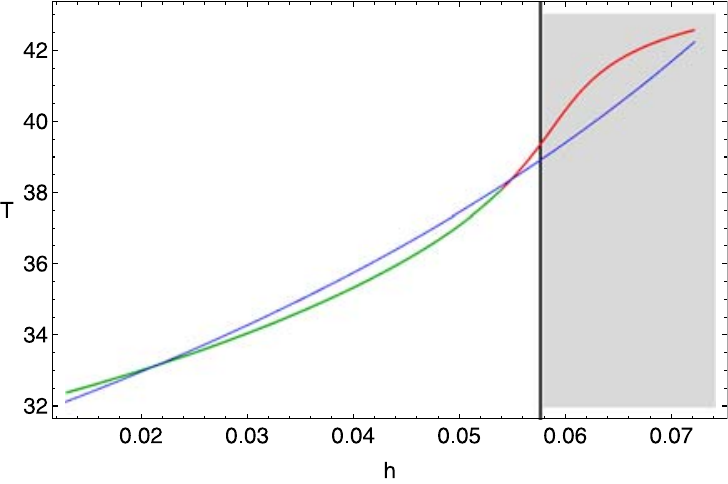}}}
\caption{Evolution of the period $T$ of the F34S family as a function of the energy $h$.}
\label{T34Sevol}
\end{figure}

\begin{figure}[!tH]
\centering
\resizebox{\hsize}{!}{\rotatebox{0}{\includegraphics*{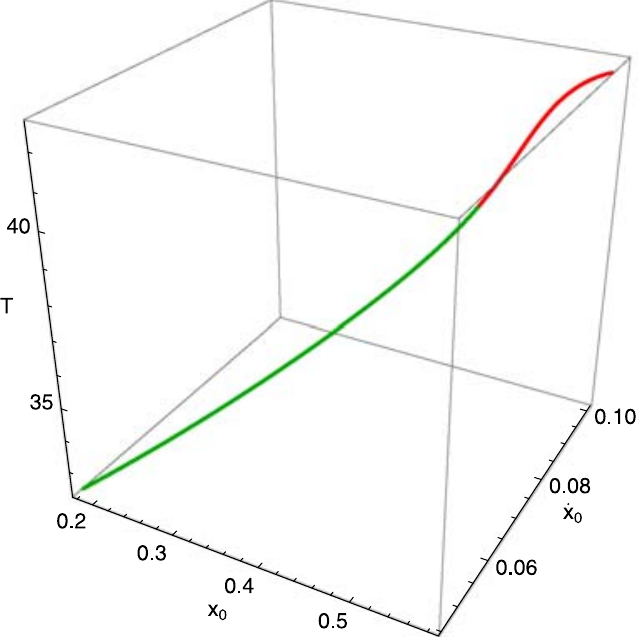}}}
\caption{Evolution of the period $T$ of the F34S family as a function of the starting position $(x_0, \dot{x_0})$ of the periodic orbits.}
\label{T34S3d}
\end{figure}

\begin{table*}
\centering
\caption{Position and period for five sample 3:4 periodic orbits of the F34S family.}
\begin{tabular}{|c||c|c|c|c|c|c|c|c|c|}
\hline
$h$  & $x_s$  & $x_f$  & $x_n$  & $\dot{x_s}$  & $\dot{x_f}$  & $\dot{x_n}$  & $T_s$  & $T_f$  & $T_n$ \\
\hline \hline
0.025 & 0.267840 & 0.265223 & 0.264454 & 0.065085 & 0.064602 & 0.064676 & 33.613031 & 33.614546 & 33.512977 \\
\hline
0.035 & 0.330003 & 0.325570 & 0.326596 & 0.080190 & 0.078255 & 0.078191 & 35.001433 & 34.501548 & 34.663479 \\
\hline
0.045 & 0.391037 & 0.389053 & 0.389344 & 0.095022 & 0.089863 & 0.089761 & 36.577436 & 36.133402 & 36.132910 \\
\hline
0.055 & 0.453701 & 0.459873 & 0.457587 & 0.110249 & 0.098133 & 0.098423 & 38.387489 & 38.680785 & 38.406571 \\
\hline
0.065 & 0.520315 & 0.535164 & 0.537571 & 0.126436 & 0.100521 & 0.100079 & 40.495923 & 41.479949 & 41.718735 \\
\hline
\end{tabular}
\label{res34S}
\end{table*}

The evolution of the F34S family is presented in Fig. \ref{res34Sevol}(a-b). We see that the value of the coordinate $x_0$ increases as we increase the energy. On the other hand, the velocity $\dot{x_0}$ initially increases but upon approaching the escape energy it starts to decrease. As expected, the majority of the computed periodic orbits are stable. However, when $h > 0.05407$ that is just before reaching the escape energy $(h_{esc} = 0.0576)$ the orbits become unstable. The escape energy, once more, cannot stop the evolution of the F34S family which ends inside the shaded area when $h = 0.07204$. The semi-numerical relation giving the initial coordinate $x_0$ of the periodic orbits can provide reliable results for the full range of the family. On the contrary, the semi-numerical relation of the initial velocity $\dot{x_0}$ is valid only when $h < 0.035$. In order to achieve much better agreement we should use the fourth order polynomial fitting curves. The terms of the fitting curves of the polynomials $x_f(h)$ and $\dot{x_f}(h)$ are given at the end of this Section in Tables (\ref{fitx}) and (\ref{fitpx}) respectively. In Fig. \ref{res34S3d} we have combined both semi-numerical Eqs. (\ref{res34Stheor}) in parametric form as a function of the energy $h$ in order to depict in three dimensions the evolution of the F34S family.

The evolution of the period of the F34S family is shown in Fig. \ref{T34Sevol}. The blue line corresponds to the semi-numerical formula (\ref{Ts}). In this case, we may say that the agreement is sufficient for the entire range of the family. However, if someone wants much more accurate results he should use the fourth order polynomial fitting curve. Fig. \ref{T34S3d} depicts an alternative approach regarding the evolution of the period of the F34S family. Here, the period is not given as a function of the energy $h$ but as a function of the starting position of the orbits using the initial coordinates $x_0$ and velocities $\dot{x_0}$ in a three-dimensional plot. Table \ref{res34S} gives the position and the period of the F34S family for five values of the energy $h$. The terms of the fourth order polynomial fitting curves giving the position and the period of the orbits of the F34S family are given in Tables (\ref{fitx}), (\ref{fitpx}) and (\ref{fitT}).

\begin{figure*}[!tH]
\centering
\resizebox{\hsize}{!}{\rotatebox{0}{\includegraphics*{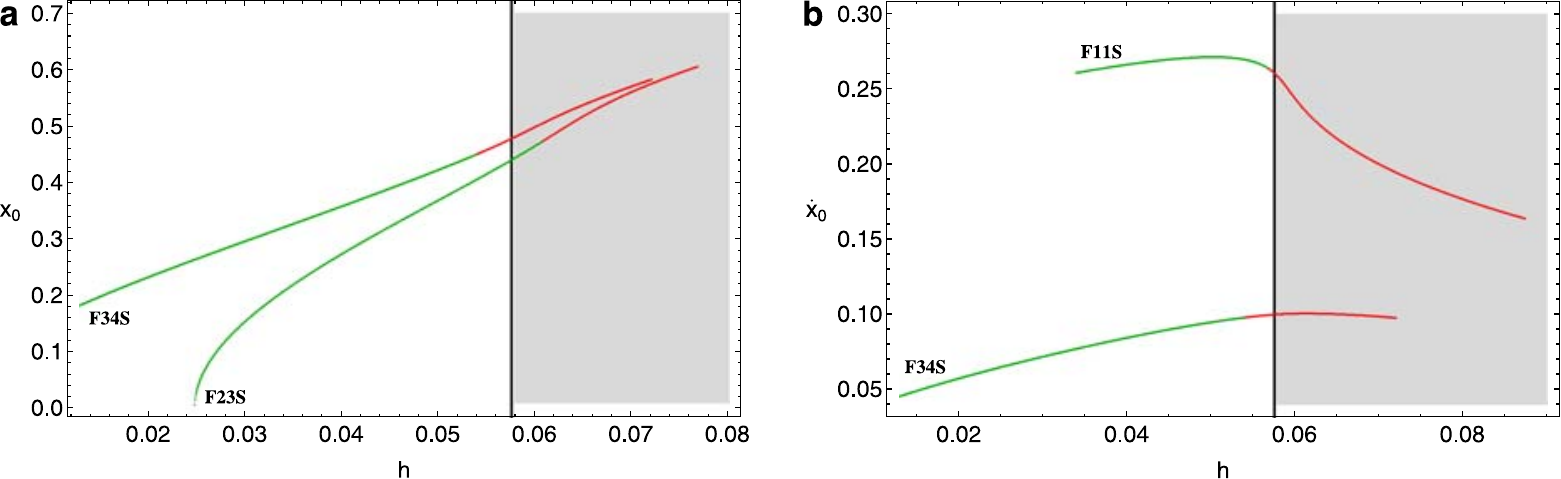}}}
\caption{(a-b): Evolution of the starting position (a-left): $x_0$ and (b-right): $\dot{x_0}$ of the periodic orbits of the families F11S, F23S and F34S as a function of the energy $h$.}
\label{res34all}
\end{figure*}

\begin{figure}[!tH]
\centering
\resizebox{\hsize}{!}{\rotatebox{0}{\includegraphics*{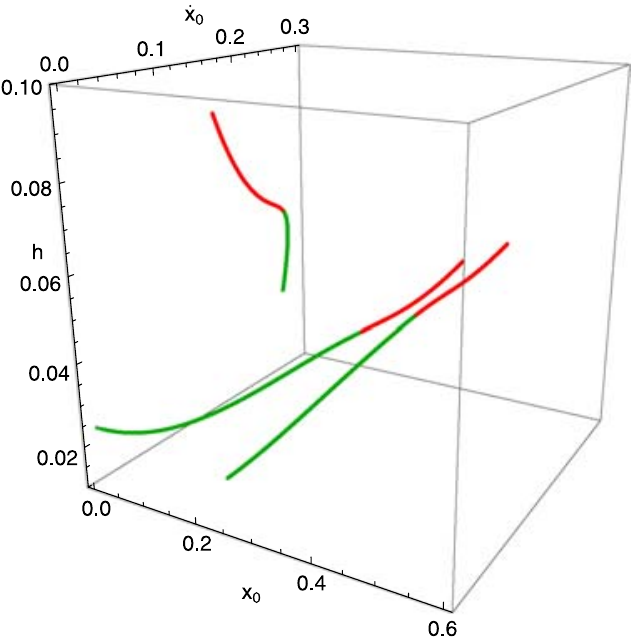}}}
\caption{Three dimensional (3D) evolution of the families F11S, F23S and F34S of periodic orbits as a function of the starting position $(x_0, \dot{x_0})$.}
\label{res343dall}
\end{figure}

\begin{figure}[!tH]
\centering
\resizebox{\hsize}{!}{\rotatebox{0}{\includegraphics*{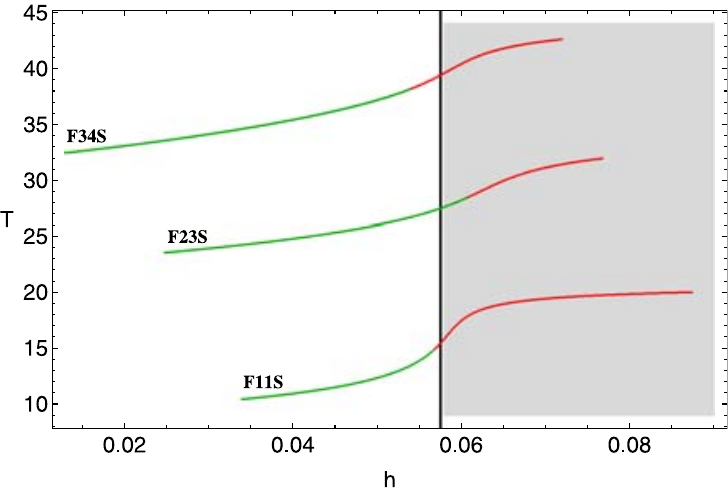}}}
\caption{Evolution of the period $T$ as a function of the energy $h$ for the families F11S, F23S and F34S.}
\label{T34all}
\end{figure}

\begin{figure}[!tH]
\centering
\resizebox{\hsize}{!}{\rotatebox{0}{\includegraphics*{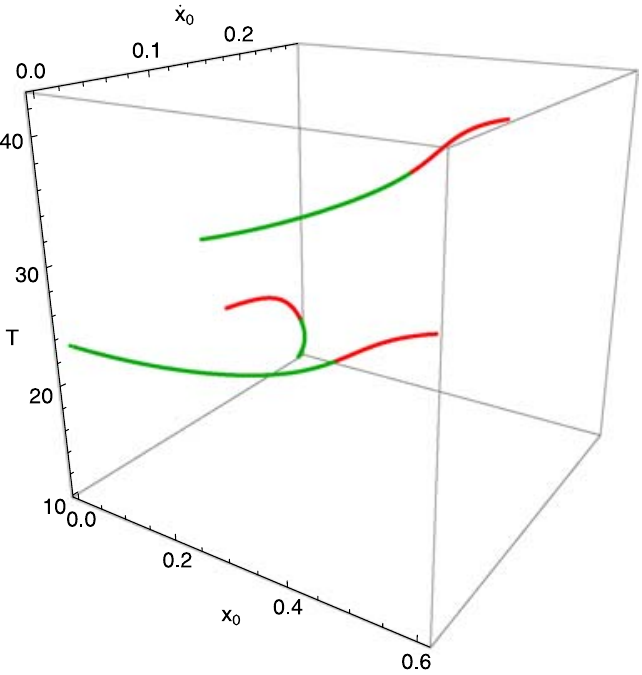}}}
\caption{Evolution of the period $T$ as a function of the starting position $(x_0, \dot{x_0})$ for the families F11S, F23S and F34S.}
\label{T343dall}
\end{figure}

In Fig. \ref{pss34} we have seen that apart from the expected 3:4 resonance, there are also two more types of resonances, that is the 1:1 and the 2:3 resonance. For these two additional types of resonances, we can locate at the $(x, \dot{x})$ phase plane the elliptic points which correspond to stable periodic orbits and also the saddle points which produce unstable periodic orbits. Thus, it would be very interesting to study the evolution of these additional resonances and present aggregated plots depicting the entire network of families of periodic orbits when the value of the energy is $h = 0.042$. We shall investigate the evolution of families which are composed of periodic orbits corresponding to elliptic points only. The two additional families are the F11S which correspond to the 1:1 resonance and the F23S family which contains the periodic orbits of the 2:3 resonance. As we can see from Fig. \ref{pss34} in order to monitor the evolution of the F11S family we have to compute each time only the initial velocity $\dot{x_0}$ of the periodic orbits. On the other hand, for the evolution of the F23S family we have to calculate the initial coordinate $x_0$.

Fig. \ref{res34all}(a-b) presents the evolution of all three families of periodic orbits in the planes $(h, x_0)$ and $(h, \dot{x_0})$. We can proceed one step further combining the two-dimensional plots presented in Fig. \ref{res34all}(a-b) in order to visualize together the evolution of all the families of periodic orbits (see Fig. \ref{res343dall}). Here we must point out, that the evolution of both families F11S and F23S is in fact two-dimensional. The evolution of the period $T$ of the three families of periodic orbits as a function of the energy $h$ is given in Fig. \ref{T34all}, while in Fig. \ref{T343dall} we present a three-dimensional plot which shows the evolution of the period as a function of the starting position $(x_0, \dot{x_0})$ of the periodic orbits. Similar results arise if we study the evolution of periodic orbits which correspond to saddle points. However, for saving space we are not going to present them here.

\begin{figure}[!tH]
\centering
\resizebox{\hsize}{!}{\rotatebox{0}{\includegraphics*{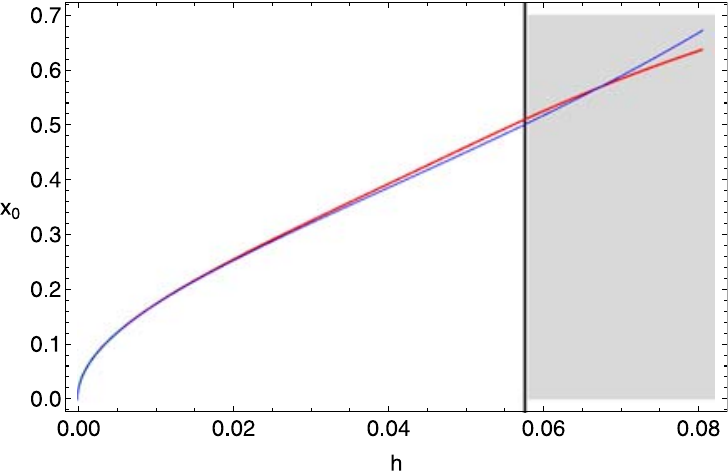}}}
\caption{Evolution of the starting position $x_0$ of the F34U family as a function of the energy $h$.}
\label{res34Uevol}
\end{figure}

\begin{figure}[!tH]
\centering
\resizebox{\hsize}{!}{\rotatebox{0}{\includegraphics*{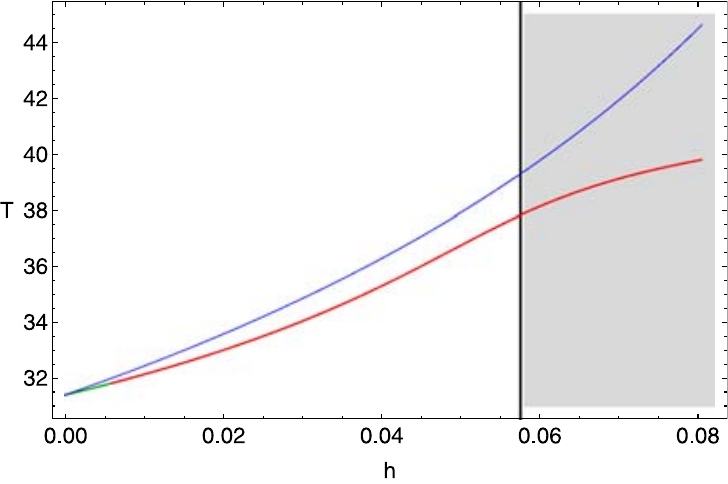}}}
\caption{Evolution of the period $T$ of the F34U family as a function of the energy $h$.}
\label{T34Uevol}
\end{figure}

We close this section, presenting the evolution of the unstable 3:4 resonant periodic orbits which compose the F34U family. Fig. \ref{res34Uevol} shows the evolution of the F34U family. It is clear, that the value of the starting position $x_0$ increases with increasing energy, while the vast majority of the computed periodic orbits are unstable. However, with a closer look at the diagram we observe that for small values of the energy $(h < 0.0059)$ that is when we are very close to the lower limit of the family, the orbits become stable. Again, the energy of escape is not able to stop the growth of the F34U family, which crosses the $h_{esc}$ barrier and ends only when $h = 0.08043$.

\begin{table*}
\centering
\caption{Position and period for five sample 3:4 periodic orbits of the F34U family.}
\begin{tabular}{|c||c|c|c|c|c|c|}
\hline
$h$  & $x_s$  & $x_f$  & $x_n$  & $T_s$  & $T_f$  & $T_n$ \\
\hline \hline
0.010 & 0.169517 & 0.172146 & 0.172827 & 32.448765 & 32.146536 & 32.152078 \\
\hline
0.025 & 0.286949 & 0.292535 & 0.289993 & 34.208754 & 33.506849 & 33.512223 \\
\hline
0.040 & 0.385050 & 0.391164 & 0.391915 & 36.290168 & 35.338257 & 35.306988 \\
\hline
0.055 & 0.482794 & 0.490943 & 0.492508 & 38.804493 & 37.430492 & 37.480336 \\
\hline
0.070 & 0.588481 & 0.585944 & 0.583402 & 41.926141 & 39.173972 & 39.118924 \\
\hline
\end{tabular}
\label{res34U}
\end{table*}

Taking into account that the periodic orbits of the F34U family start perpendicularly from the $x$-axis, the starting position $x_0$ can be obtained using Eq. (\ref{xs}). Our numerical calculations suggest, that in order to achieve the best agreement between theoretical and numerical outcomes, the correction term should be $c = \sqrt{n}/2\omega_1$. Thus, combining Eqs. (\ref{freq}) and (\ref{xs}) and taking into account that $\omega_2 = 4 \omega_1/3$ the relation for the starting position of periodic orbits of the F34U family becomes
\begin{equation}
x_s = 4\omega_1 \sqrt{\frac{2\omega_1 h}{2m^2\omega_1^5 - n^2\sqrt{3}\varepsilon h}},
\label{xs34U}
\end{equation}
where $n = 3$ and $m = 4$. The plot of Eq. (\ref{xs34U}) is shown as the blue line in Fig. \ref{res34Uevol}. In this case, the semi-numerical formula (\ref{xs34U}) can approximate with sufficient accuracy the evolution of the entire F34U family of periodic orbits.

\begin{table*}
\centering
\caption{Terms of the fourth order polynomial fitting curves for the coordinate $x_0$ for each resonance case.}
\begin{tabular}{|c||c|c|c|c|c|}
\hline
Family & $x_0$ & $x_1$ & $x_2$ & $x_3$ & $x_4$ \\
\hline \hline
F11U & 0.046155 & 33.7568 & -1458.92 & 52486.1 & -717946 \\
\hline
F12S & 0.033446 & 22.4516 & -886.436 & 13192.8 & 216030 \\
\hline
F12U & 0.065187 & 23.9601 & -554.035 & 11233 & -84457.9 \\
\hline
F13S & - & - & - & - & - \\
\hline
F13U & 0.073318 & 21.28 & 424.165 & 5970.85 & -28746.6 \\
\hline
F23S & 0.033167 & 46.584 & -4199.97 & 245129 & -4.58$\times 10^{6}$ \\
\hline
F23U & - & - & - & - & - \\
\hline
F34S & -0.05187 & 25.2923 & -764.302 & 12429.1 & -69414.2 \\
\hline
F34U & 0.047099 & 14.7885 & -298.362 & 4628.22 & -25900 \\
\hline
\end{tabular}
\label{fitx}
\end{table*}
\begin{table*}
\centering
\caption{Terms of the fourth order polynomial fitting curves for the velocity $\dot{x_0}$ for each resonance case.}
\begin{tabular}{|c||c|c|c|c|c|}
\hline
Family & $\dot{x_0}$ & $\dot{x_1}$ & $\dot{x_2}$ & $\dot{x_3}$ & $\dot{x_4}$ \\
\hline \hline
F11U & - & - & - & - & - \\
\hline
F12S & 0.012982 & 9.34655 & -428.818 & 10344.3 & -27110.1 \\
\hline
F12U & - & - & - & - & - \\
\hline
F13S & 0.027768 & 8.66284 & -182.744 & 1822.13 & -993038 \\
\hline
F13U & - & - & - & - & - \\
\hline
F23S & - & - & - & - & - \\
\hline
F23U & 0.016655 & 14.3045 & -861.263 & 21286.6 & -236690 \\
\hline
F34S & -0.00972 & 6.25877 & -200.843 & 3437.72 & -21909.1 \\
\hline
F34U & - & - & - & - & - \\
\hline
\end{tabular}
\label{fitpx}
\end{table*}
\begin{table*}
\centering
\caption{Terms of the fourth order polynomial fitting curves for the period $T$ for each resonance case.}
\begin{tabular}{|c||c|c|c|c|c|}
\hline
Family & $T_0$ & $T_1$ & $T_2$ & $T_3$ & $T_4$ \\
\hline \hline
F11U & 15.7523 & 115.183 & 10391.6 & -324319 & 2.35$\times 10^{6}$ \\
\hline
F12S & 15.8641 & -103.212 & 45391.8 & -3.515$\times 10^{6}$ & 9.275$\times 10^{7}$ \\
\hline
F12U & 15.8013 & 40.9863 & 3640.15 & -30221.3 & -134665 \\
\hline
F13S & 15.7031 & 20.3504 & 2808.94 & -114887 & 1.583$\times 10^{6}$ \\
\hline
F13U & 15.6129 & 63.5187 & -1750.8 & 47919.4 & -291236 \\
\hline
F23S & 31.5849 & -41.426 & 135928 & -1.747$\times 10^{7}$ & 8.105$\times 10^{8}$ \\
\hline
F23U & 31.4894 & 219.212 & 11486.6 & 1.117$\times 10^{6}$ & -5.071$\times 10^{7}$ \\
\hline
F34S & 31.1037 & 171.971 & -6894.02 & 175569 & -1.121$\times 10^{6}$ \\
\hline
F34U & 31.4245 & 70.0258 & -72.3384 & 32341.7 & -328656 \\
\hline
\end{tabular}
\label{fitT}
\end{table*}

In Fig. \ref{T34Uevol} we provide a plot depicting the evolution of the period of the F34U family. The blue line corresponds to the semi-numerical formula (\ref{Ts}). It is evident, that the semi-numerical formula fails to approximate almost the entire family. Thus, the fourth order polynomial fitting curve is the only solution in order to obtain semi-numerically the period of the orbits which belong to the F34U family. In Table \ref{res34U} we provide the position and the period of the F34U family for five sample values of the energy $h$.

\section{Escaping orbits}
\label{Escape}

\begin{figure}[!tH]
\centering
\resizebox{\hsize}{!}{\rotatebox{0}{\includegraphics*{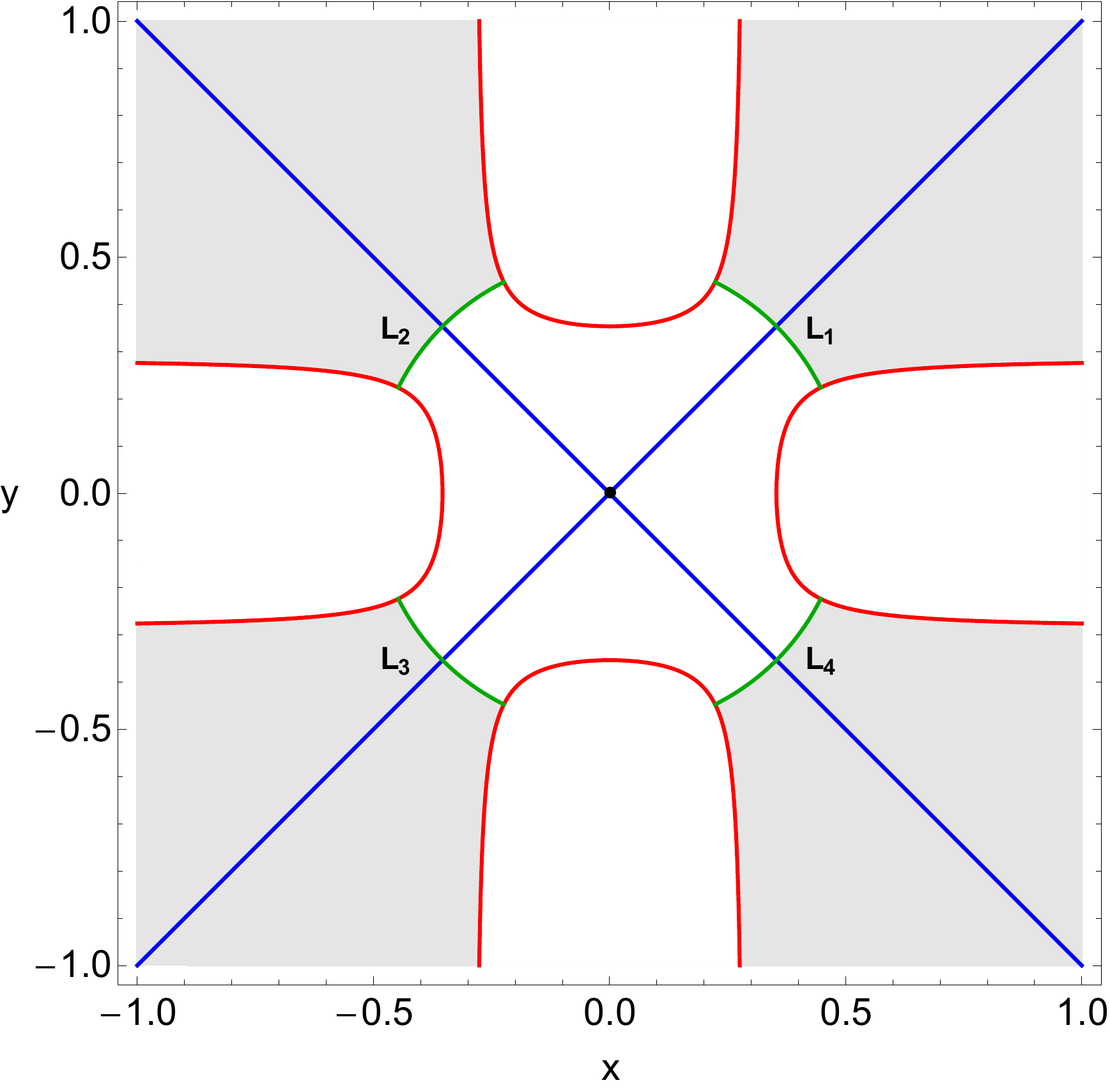}}}
\caption{The open ZVC when $\omega_1 = \omega_2 = 0.4$ and $h = 0.01$. $L_1$, $L_2$, $L_3$ and $L_4$ indicate the four unstable Lyapunov orbits plotted in green, while with blue color we present the straight-lines $y = \pm \omega_1/\omega_2 x$.}
\label{EscLyap}
\end{figure}

In the previous Section, we investigated the evolution of several families of resonant periodic orbits. We found that apart from the F11S family which consist of the 1:1 straight-line periodic orbits, all the other families continue to exist beyond the escape energy. Moreover, our experiments indicate that the vast majority of the computed periodic orbits are unstable for values of energy larger than the escape energy $h_{esc}$.

\begin{figure}[!tH]
\centering
\resizebox{\hsize}{!}{\rotatebox{0}{\includegraphics*{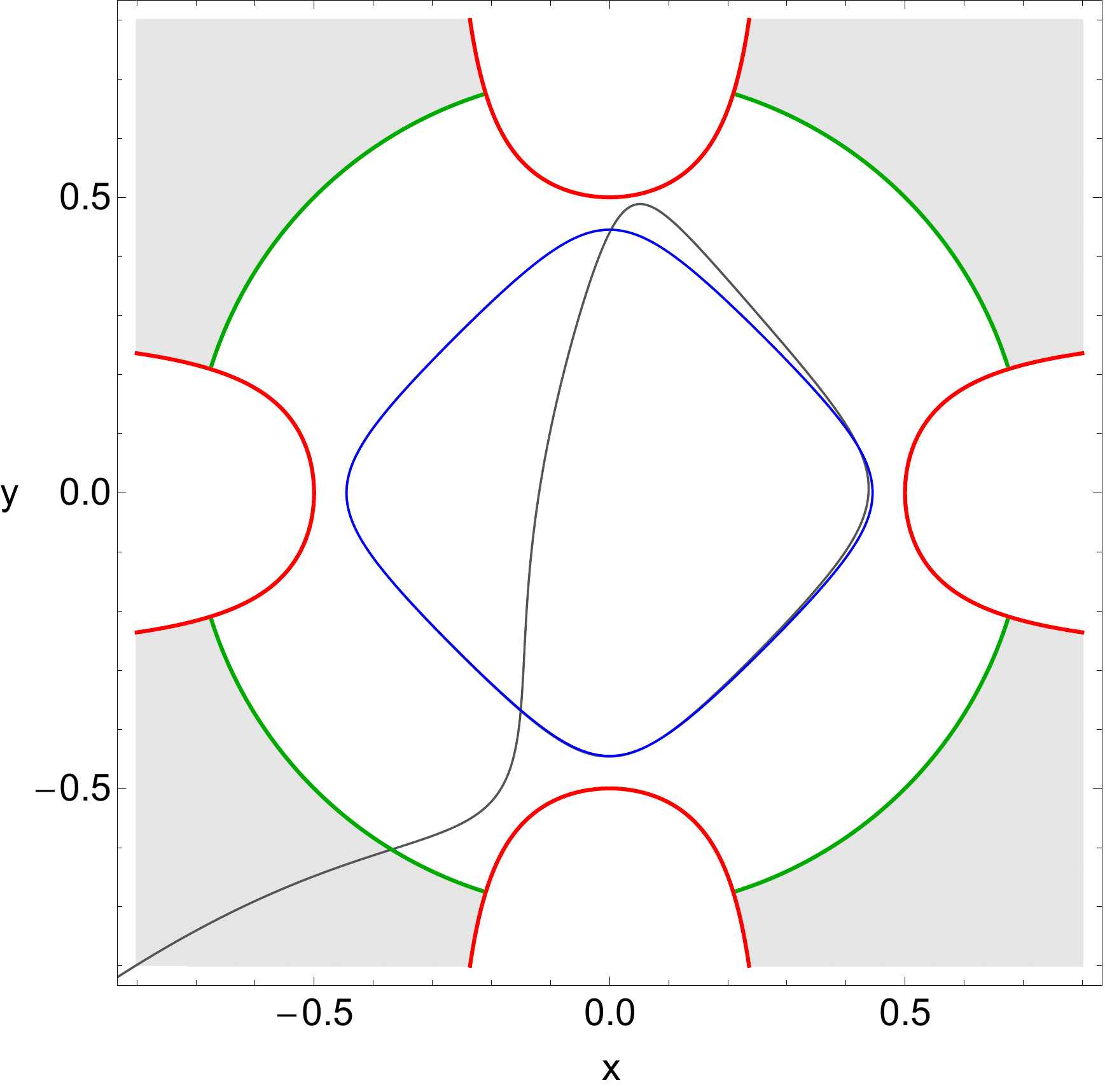}}}
\caption{An escaping orbit which belongs to the F11U family.}
\label{orb11U_esc}
\end{figure}

\begin{figure}[!tH]
\centering
\resizebox{\hsize}{!}{\rotatebox{0}{\includegraphics*{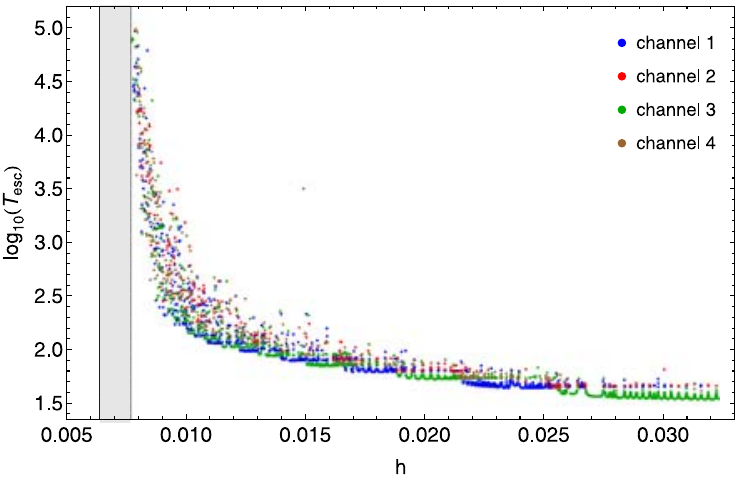}}}
\caption{Evolution of the escape period $T_{esc}$ of the orbits composing the F11U family.}
\label{F11U_Tesc}
\end{figure}

\begin{figure}[!tH]
\centering
\resizebox{\hsize}{!}{\rotatebox{0}{\includegraphics*{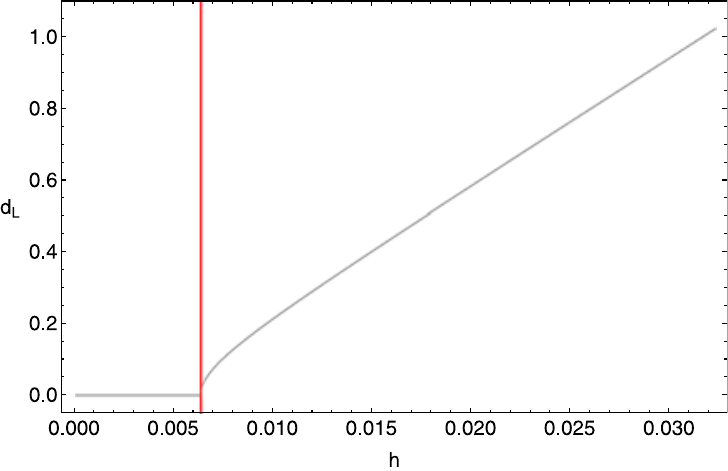}}}
\caption{Evolution of $d_L$ as a function of the energy $h$.}
\label{lencha}
\end{figure}

\begin{figure}[!tH]
\centering
\resizebox{\hsize}{!}{\rotatebox{0}{\includegraphics*{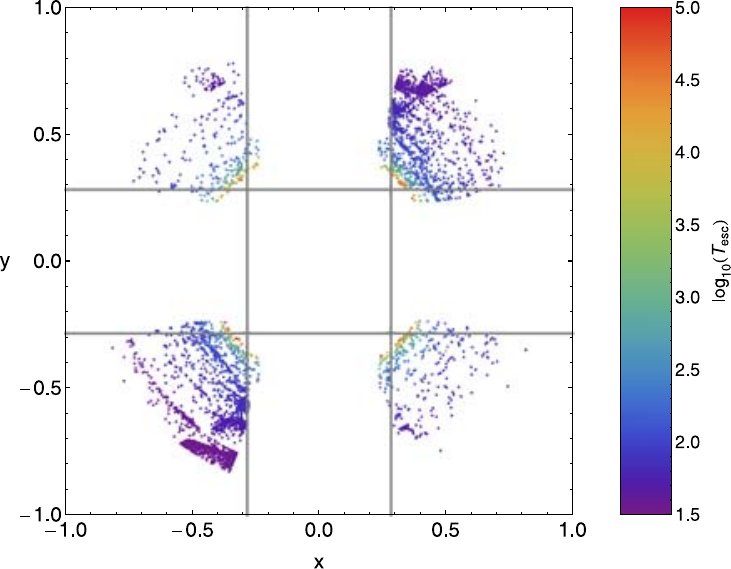}}}
\caption{Relation between escaping position and escaping period for the F11U family. The gray curve corresponds to the equipotential of the escape energy $h_{esc}$.}
\label{F11U_chas}
\end{figure}

\begin{figure}[!tH]
\centering
\resizebox{\hsize}{!}{\rotatebox{0}{\includegraphics*{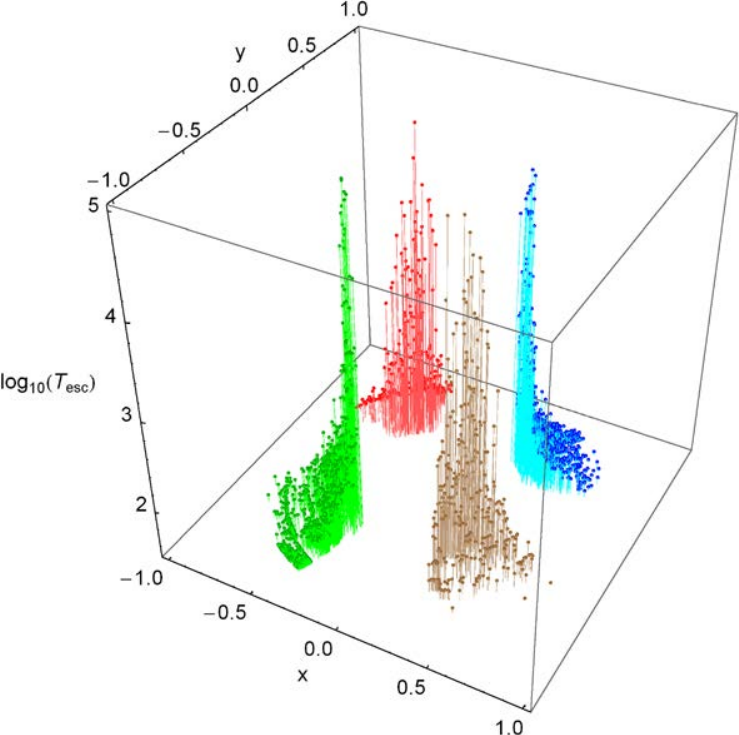}}}
\caption{A three-dimensional plot depicting the connection between the escaping position and escaping period for the F11U family.}
\label{cha3D}
\end{figure}

When the value of the energy $h$ is smaller than the escape energy, potential (\ref{pot}) has closed Zero Velocity Curves (ZVCs). On the other hand, when $h > h_{esc}$ the equipotential curves are open and extend to infinity. An open ZVC consists of four branches forming four channels through which an orbit can escape to infinity. Therefore, we have four openings symmetric with respect to the $x$ and $y$ axis. At every opening there is an unstable periodic orbit close to the line of maximum potential [\citealp{11}] which is called a Lyapunov orbit. Such an orbit reaches the ZVC, on both sides of the opening crossing perpendicularly one of the lines $y = \pm \omega_1/\omega_2 x$ and returns along the same path thus, connecting two opposite branches of the ZVC. Lyapunov orbits are very important for the escapes from the system, since if an orbit intersects any one of these orbits with velocity outwards moves always outwards and eventually escapes from the system without any further intersections with the surface of section. The open ZVC when $\omega_1 = \omega_2 = 0.4$ and $h = 0.01 > h_{esc}$ is presented with red color in Fig. \ref{EscLyap}. In the same plot we denote the four Lyapunov orbits by $L_1$, $L_2$, $L_3$ and $L_4$ using green color, while the straight-lines $y = \pm \omega_1/\omega_2 x$ are plotted in blue.

In this Section, we shall study the parts of the several resonant families of periodic orbits for which $h > h_{esc}$. When $h > h_{esc}$ the periodic orbits remain or become unstable. If we combine this result with the fact that for energies higher than $h_{esc}$ the ZVC opens creating four escape channels then, it is natural to assume that most, if not all, the unstable periodic orbits will escape. Therefore, our main objective is to calculate the escape period of the orbits. The escape period is defined as the required time interval so that an orbit to intersect one of the four Lyapunov orbits. Since there are four possible symmetric channels of escape, there is no obvious reason to assume that all orbits will escape from the same channel. So, apart from the escape period we shall also focus on the different channels of escape. We name each channel according to the corresponding Lyapunov orbit. Therefore, $L_1$ orbit indicates escape channel 1 (upper right), $L_2$ orbit indicates escape channel 2 (upper left), $L_3$ orbit indicates escape channel 3 (lower left) and $L_4$ orbit indicates escape channel 4 (lower right) (see Fig. \ref{EscLyap}). In every resonance case, we defined the appropriate increment value regarding the energy so that the interval $(h_{esc}, h_{max}]$ contains always 5000 periodic orbits ($h_{max}$ is the maximum value of the energy for which a resonance exists, or in other words, the upper limit of each resonance family). For the numerical integration of orbits, we set a maximum time of $10^5$ time units. If an orbit does not intersect any of the four Lyapunov orbits within this time interval, then we consider it as trapped.

\begin{table*}
\centering
\caption{The initial conditions, the period, the escape period and the energy for the escaping orbits shown in Fig. \ref{EscOrbs}. In all cases $y_0 = 0$, while the value of $\dot{y_0}$ is always obtained from the energy integral (\ref{ham}).}
\begin{tabular}{|c||c|c|c|c|c|c|}
\hline
Figure & h & $x_0$ & $\dot{x_0}$ & $T_{per}$ & $T_{esc}$ & channel \\
\hline \hline
\ref{EscOrbs}a & 0.027 & 0.37367874 & 0.14386698 & 27.50824362 & 60.44183494 & 1 \\
\hline
\ref{EscOrbs}b & 0.050 & 0.75165858 & 0.00000000 & 22.30059423 & 45.24039298 & 3 \\
\hline
\ref{EscOrbs}c & 0.065 & 0.00000000 & 0.30503433 & 25.98205420 & 57.19004295 & 4 \\
\hline
\ref{EscOrbs}d & 0.080 & 0.93830538 & 0.00000000 & 22.11514906 & 57.86284495 & 3 \\
\hline
\ref{EscOrbs}e & 0.016 & 0.42051683 & 0.00000000 & 46.13014861 & 217.13285611 & 1 \\
\hline
\ref{EscOrbs}f & 0.020 & 0.00000000 & 0.08976888 & 41.20466209 & 116.47449880 & 4 \\
\hline
\ref{EscOrbs}g & 0.065 & 0.53757159 & 0.10007926 & 41.71873582 & 100.79256284 & 4 \\
\hline
\ref{EscOrbs}h & 0.070 & 0.58340220 & 0.00000000 & 39.11892461 & 86.17466687 & 3 \\
\hline
\end{tabular}
\label{esc_orbs}
\end{table*}

\begin{table}
\centering
\caption{Analytical data about trapped and escaping orbits in every resonance case.}
\setlength{\tabcolsep}{4.0pt}
\begin{tabular}{|c||c|c|c|c|c|}
\hline
Family & Trapped & channel 1 & channel 2 & channel 3 & channel 4 \\
\hline \hline
F11U & 277 & 1736 & 296 & 2361 & 330 \\
\hline
F12S & 0 & 2875 & 961 & 503 & 661 \\
\hline
F12U & 0 & 2012 & 2080 & 515 & 393 \\
\hline
F13S & 760 & 522 & 870 & 495 & 2353 \\
\hline
F13U & 0 & 2081 & 645 & 1636 & 638 \\
\hline
F23S & 0 & 703 & 874 & 2528 & 895 \\
\hline
F23U & 0 & 783 & 1503 & 1992 & 722 \\
\hline
F34S & 0 & 778 & 888 & 2108 & 1226 \\
\hline
F34U & 0 & 1291 & 786 & 1728 & 1195 \\
\hline
\end{tabular}
\label{chans}
\end{table}

We already proved that the F11S family which consists of the 1:1 straight-line periodic orbits ends when $h = h_{esc}$ and therefore, in this case there are no escaping orbits. Thus, we begin our investigation with the F11U family. Fig. \ref{orb11U_esc} shows an escaping 1:1 resonant periodic orbit. The initial conditions of this orbit are $x_0 = 0.44511719$ and $y_0 = \dot{x_0} = 0$, while the value of the energy is $h = 0.02$. This orbit is plotted with blue color until $t = T$, where $T = 20.01103827$ is the period of the orbit. For $t > T$ the orbit is plotted with gray color. The escape period of this orbit is $T_{esc} = 53.74725496$. When $t = T_{esc} \simeq 2.7 T$ the orbit intersects the $L_3$ Lyapunov orbit and escapes from channel 3. In Fig. \ref{F11U_Tesc} we present the escape period for all the orbits of the F11U family. Each dot represents an orbit of the F11U family. Moreover, we used four different colors to indicate through which channel each orbit escapes. In the same plot we observe, that for $h_{esc} \leq h \leq 0.00765$ there is a gray shaded area. This energy range contains 277 unstable periodic orbits that do not escape even if we integrated them for $10^5$ time units. Perhaps, they require larger integration time to escape. However, in our investigation these orbits are considered trapped.

It is evident from Fig. \ref{F11U_Tesc} that the escape period of the orbits is decreasing rapidly as the value of the energy increases. We feel it is important to justify this phenomenon. As previously explained, a Lyapunov orbit acts like a barrier, reaches the ZVC on both sides of the opening and returns along the same path thus, connecting two opposite branches of the ZVC. Therefore, we could exploit this fact in order to quantify the escape phenomenon. For this purpose, we define $d_L$ as the distance between the two points at which the Lyapunov orbit touches the two opposite branches of the ZVC. Using $d_L$ we can measure, in a way, the width of the escape channel. Fig. \ref{lencha} depicts the evolution of $d_L$ as a function of the energy $h$. The red vertical line corresponds to the escape energy. Obviously, when $h < h_{esc}$ $d_L = 0$ since the ZVC is close. On the other hand, when $h > h_{esc}$ $d_L$ increases almost linearly. This means that as we increase the value of the energy the escape channels becoming more wide and therefore, the unstable orbits need less and less time in order to find one of the four openings and eventually escape from the system. Thus, we may say that we justified the decreasing pattern of $T_{esc}$ shown in Fig. \ref{F11U_Tesc}.

\begin{figure*}
\centering
\resizebox{0.8\hsize}{!}{\rotatebox{0}{\includegraphics*{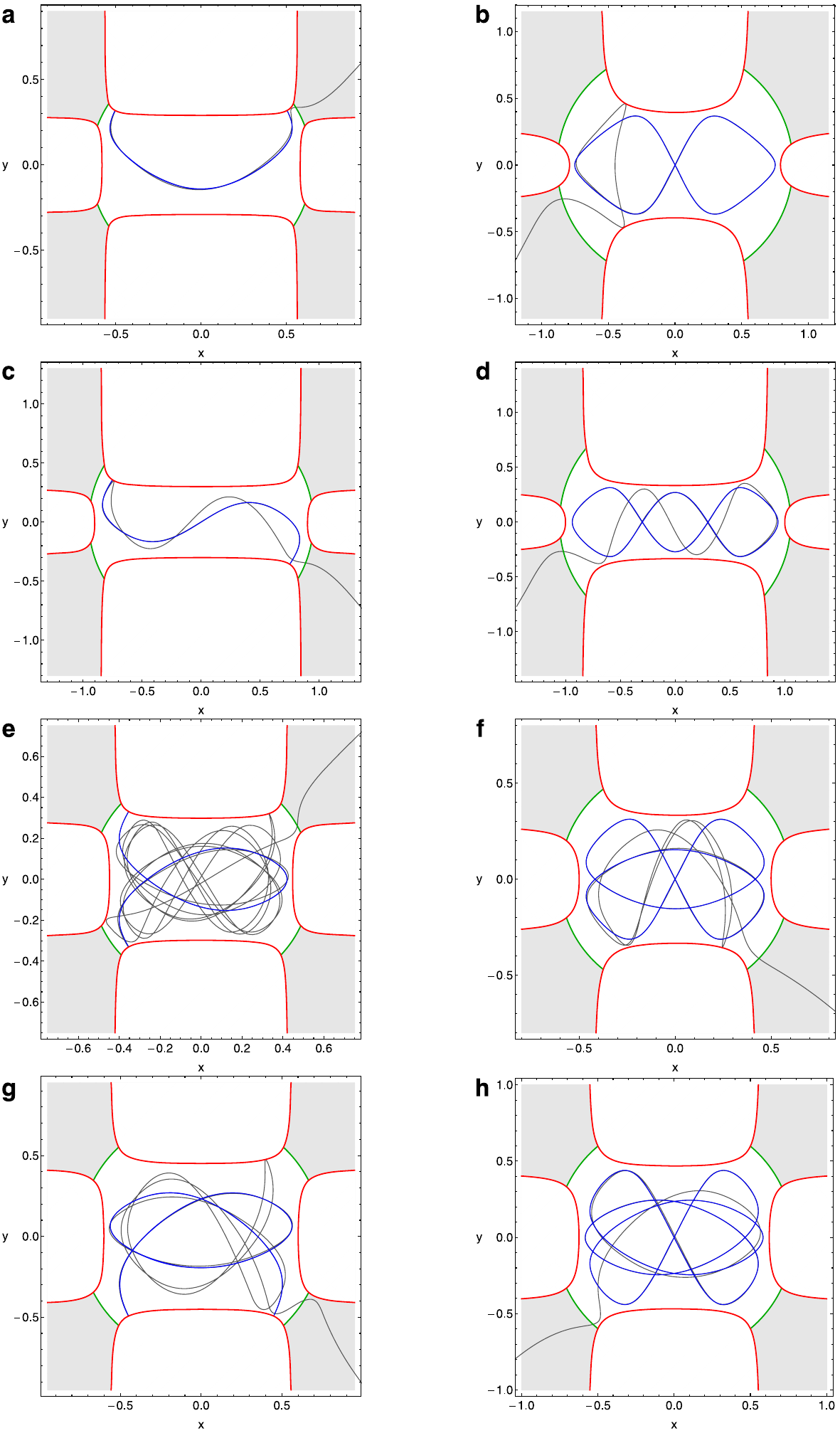}}}
\caption{(a-h): Representative examples of escaping orbits in each resonance case. More information about these orbits are given in Table \ref{esc_orbs}.}
\label{EscOrbs}
\end{figure*}

\begin{figure*}
\centering
\resizebox{\hsize}{!}{\rotatebox{0}{\includegraphics*{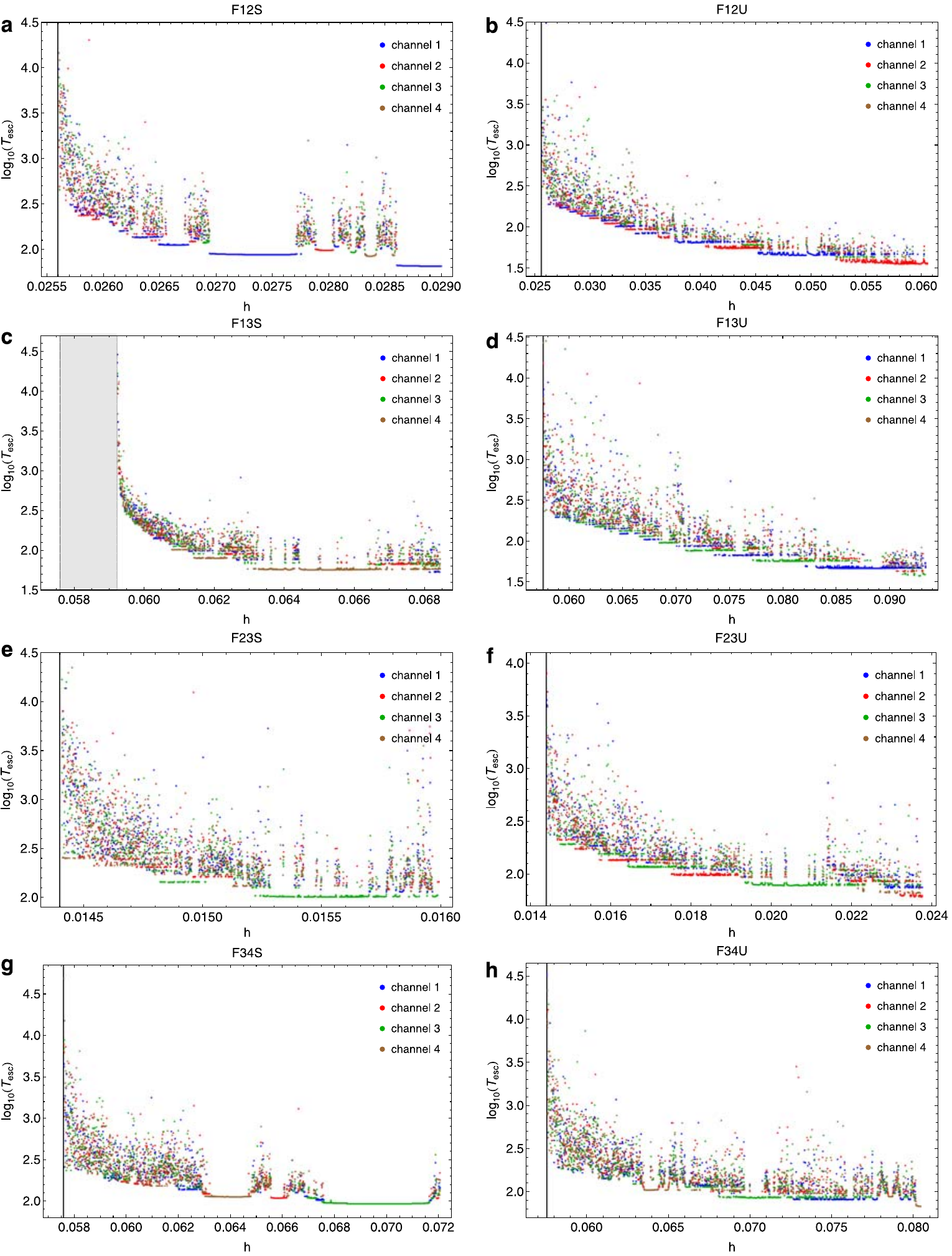}}}
\caption{(a-h): Evolution of the escape period $T_{esc}$ in every resonance case.}
\label{ResTesc}
\end{figure*}

\begin{figure*}
\centering
\resizebox{0.8\hsize}{!}{\rotatebox{0}{\includegraphics*{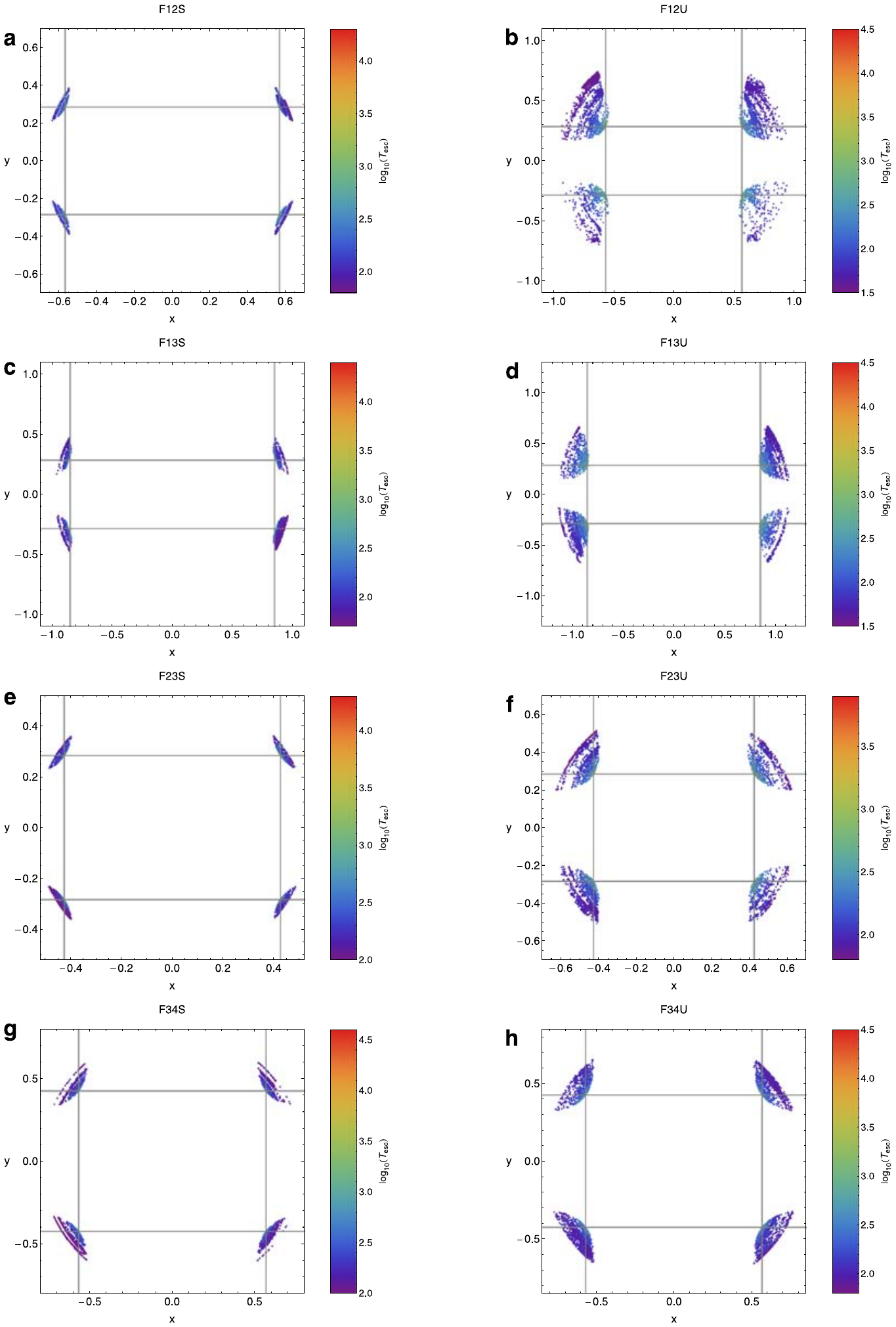}}}
\caption{(a-h): Connection between the escaping position and the escaping period of the orbits in every resonance case.}
\label{ResChans}
\end{figure*}

It would be of particular interest to locate the escape positions of the orbits. By the term ``escape positions" we refer to the points on the $(x,y)$ plane at which the orbits intersect one of the Lyapunov orbits. Fig. \ref{F11U_chas} depicts the escape positions of the F11U family. Each dot corresponds to an escaping orbit and is plotted with different color according to the escape period of each orbit. The relation between the range of the colors and the corresponding escape periods is given in the color bar at the right part of the plot. The same quantities that is the escape position and the escape period, are presented in the three-dimensional plot of Fig. \ref{cha3D}. Here, we use four different colors according to Fig. \ref{F11U_Tesc} in order to illustrate the four different escape channels. We observe, that as we approach to values of energy very close to $h_{esc}$ the escape period is growing rapidly.

Following the same philosophy as above, we can study the escaping orbits in all the resonance cases. In Fig. \ref{EscOrbs}(a-h) we present representative examples of escaping orbits in each resonance case. The initial conditions, the energy, the period and also the escape period of the orbits are provided in Table \ref{esc_orbs}. Fig. \ref{ResTesc}(a-h) shows the evolution of the escape period in all resonance cases. Moreover, in the same plot we can distinguish between the four possible channels of escape. Of particular interest is the plot shown in Fig. \ref{ResTesc}c, that is the case of the F13S family. We observe, that when $h_{esc} < h < 0.05918$ there is a gray shaded area which contains 760 orbits. These orbits are stable (see Fig. \ref{res13Sevol}) and do not escape. These orbits remained trapped even if we integrated them again using $10^6$ time units of numerical integration. Therefore, we have strong numerical evidence, that these particular periodic orbits remain trapped regardless of the total time of the numerical integration. Table \ref{chans} contains the exact number of trapped orbits and also the number of orbits per escape channel in all resonance cases. We may say, that almost in every resonance case there is a favorite channel chosen by most of orbits. Finally, in Fig. \ref{ResChans}(a-h) we can see through color scaled plots the escaping position and the corresponding escape period of the orbits in every resonance case.

\section{Discussion and conclusions}
\label{disc}

In this research, we tried to shed some light on the evolution of families of periodic orbits in a dynamical system composed of a two-dimensional perturbed harmonic oscillator. In particular, the aim of this work was not only to locate the position and compute the period of the periodic orbits but also to monitor the evolution of each resonance family as a function of the energy. The results obtained from the semi-numerical methods were always compared with the corresponding ones derived by the numerical integration of the equations of motion. The ratio of the unperturbed frequencies was equal to a rational number $n:m$, where $n \leq 3$ and $m \leq 4$. There are two main reason justifying our choice of the particular values of $n$ and $m$. First, is not feasible to to study all the possible resonance cases in a given potential and second, because the important resonance cases are those with low values of $n$ and $m$. Here we must point out, that we considered only cases where $\omega_1:\omega_2 = n:m$.

To our investigation we used the semi-numerical relations presented in Paper I. The formula giving the starting point of the periodic orbit mainly depends on the type of the orbit. In particular, there are two main types of periodic orbits. The first type corresponds to orbits starting perpendicularly from the $x$-axis. These orbits are usually stable periodic orbits which avoid going through the center and are called ``centrophobic". The second type consists usually of unstable periodic orbits starting from the origin which are known as ``centrophilic". On the other hand, the semi-numerical relation regarding the period of the periodic orbit is common for both types of orbits. Our initial experiments revealed that these semi-numerical relations cannot provide equally accurate results in all resonance cases. On this basis, we inserted a correction term which allowed us to manipulate the semi-numerical expressions separately in every resonance case in order to obtain much more reliable results. However, we found out that especially the formula giving the period of the periodic orbits in most of the studied cases is not able to follow the evolution of the entire family of orbits. This is true, because all the semi-numerical relations presented in Paper I are very simple and therefore, their accuracy is reduced significantly especially in cases where the families of the resonant periodic orbits present either turning points or a non-monotone behavior.

In an attempt to overtake this drawback, we exploited the numerical results obtained from the numerical integration. Taking into account that each resonance family is very ``rich" containing 5000 periodic orbits that have been integrated numerically, we interpolated these data thus obtaining a fourth-order polynomial fitting curve. Therefore, we defined fourth-order polynomials giving not only the period but also the position of the periodic orbits in each resonance family. By applying this method, we achieved the best possible agreement between the numerical results obtained from the numerical integration of the equations of motion and the semi-numerical outcomes from the interpolation. In fact, in every resonance case the relative error between numerical and semi-numerical results was always less than 1\%. The sufficient approximation which yields to a very small and therefore negligible relative error proves beyond any doubt the great efficiency of the semi-numerical methods. The main advantage of this procedure is that the fourth-order polynomials giving the position and the period of the periodic orbits contain only one variable which is the value of the energy.

We began our investigation from the F11S family which contains the 1:1 straight-line periodic orbits. This is indeed a very interesting family because the periodic orbits of this family undergo an infinity of transitions to instability and stability as $h$ tends to $h_{esc}$, while the period of these orbits tends to infinity. F11S family ends when $h = h_{esc}$ because for values of energy larger than the escape energy the ZVC opens and the orbits escape immediately to infinity. On the contrary, our numerical calculations indicated that all the other families of resonant periodic orbits continue to exist beyond the escape energy. Moreover, we found that the vast majority of the periodic orbits remain or become unstable when $h > h_{esc}$. We observed, that each resonance family behaves differently upon approaching the escape energy. In fact, the evolution of a resonance family beyond $h_{esc}$ depends mainly on the particular type of the periodic orbits of the family.

As we have seen, in each resonance case there are two types of periodic orbits: (i) stable periodic orbits corresponding to elliptic points in the PSS and (ii) unstable periodic orbits corresponding to saddle points in the phase plane. If we compare the periodic orbits shown in Figs. \ref{orb11}, \ref{orbs12}a, \ref{orbs13}a, \ref{orbs23}a and \ref{orbs34}a with those presented in Figs. \ref{orb11}, \ref{orbs12}b, \ref{orbs13}b, \ref{orbs23}b and \ref{orbs34}b we shall observe a very interesting phenomenon. In all cases, the stable resonant periodic orbits (green color) ``touch" the ZVC, while all the unstable periodic orbits (red color) do not interact with the ZVC. Therefore, when $h > h_{esc}$ the first type of periodic orbits which require the presence of the ZVC in order to be reflected and return along the same path, escapes much more quickly than the second type. This can also be supported by looking the evolution of each resonance family. Indeed, if we compare Figs. \ref{res11Sevol}, \ref{res12Sevol}, \ref{res13Sevol}, \ref{res23Sevol} and \ref{res34Sevol} with Figs. \ref{res11Uevol}, \ref{res12Uevol}, \ref{res13Uevol}, \ref{res23Uevol} and \ref{res34Uevol}, it becomes obvious that always the family which consists of periodic orbits of type (i) ends much more quickly than the family composed of type (ii) resonant periodic orbits.

Of particular interest, was the cases of the F12S and F34S families of periodic orbits. Studying the evolution of these particular families of resonant periodic orbits was a real challenge due to the peculiar nature of these orbits. These orbits do not pass through the origin and also do not start perpendicularly from the $x$-axis. Therefore, we could not use non of the semi-numerical relations of Paper I in order to compute their location. Thus, we had to develop new semi-numerical formulas to obtain the location of the periodic orbits.

Finally, we made a thorough investigation in all families of periodic orbits when $h > h_{esc}$. Our main objective, was to compute the escape period and also to determine the escape position of the orbits. The escape period is defined as the required time interval so that an orbit to intersect one of the four Lyapunov orbits. Our numerical calculations, suggested that the escape period in all families is rapidly reduced as the value of the energy increases. Only for values of the energy very close to $h_{esc}$ the escape time is considerable high. This can be explained, if we take into account that when the energy increases we have a simultaneous increase on the width of the escape channels. The width of an escape channel can, in a way, be defined as the distance between the two points at which a Lyapunov orbit intersects the two opposite branches of the ZVC. Thus, as the value of the energy increases the escape channels becoming more and more wide and therefore, the unstable orbits need less and less time in order to find one of the four available openings and eventually escape to infinity.

In the case of the F11U family, we found that there is a small energy range beyond $h_{esc}$ which contains several unstable 1:1 periodic orbits that do not escape even if we integrated them for a time interval of $10^5$ time units. Therefore, these are trapped unstable orbits. Considering the unstable nature of these orbits, it is very likely, that if we integrate them for much more time they will eventually escape. On the other hand, the trapped periodic orbits we found in F13S family are stable and therefore, we have strong numerical evidence that they remain trapped regardless of the total time of the numerical integration. We may say, that in general terms, the choice regarding which escape channel will follow each orbit is completely random, even though that in some resonance families we can distinguish minor patterns only for small ranges of the energy. However, almost in every resonance case there is a favorite channel chosen by the majority of the orbits.

We consider the outcomes of the present research as an initial effort in the task of exploring the evolution of families of resonant periodic orbits. Since our current results are encouraging, it is in our future plans to study and reveal the evolution and the stability of periodic orbits in more complicated dynamical system such as galactic models. Moreover, we plan to expand our numerical methods in order to trace and locate three-dimensional resonant periodic orbits.

\section*{Acknowledgments}

I would like to express my warmest thanks to the two anonymous referees for the careful reading of the manuscript and for their very positive comments and aptly suggestions, which allowed us to improve both the quality and the clarity of the present article.

\end{document}